\documentclass[journal,10pt,twocolumn]{IEEEtran}
\IEEEoverridecommandlockouts
\usepackage{cite}
\usepackage[caption=false]{subfig}
\usepackage{stfloats}
\usepackage{amsmath,amssymb,amsfonts}
\usepackage{algorithmic}
\usepackage{graphicx}
\usepackage{textcomp}
\usepackage{xcolor}
\usepackage{booktabs}
\usepackage{tikz}
\usetikzlibrary{
        calc,
        matrix,
        spy,
        shapes,
        plotmarks,
        arrows.meta,
        colorbrewer,
        backgrounds,
    }
\usepackage{blochsphere}
\usepackage{pgfplots}
\pgfplotsset{compat=1.18}
\usepackage{esvect}
\usepackage{stackengine}
\usepackage{grffile}
\usepackage{siunitx}
\usepackage{textpos}
\usepackage{hyperref}
\usepackage{transparent}
\usepackage[inline]{asymptote}
\usepackage{xparse}
\usepackage{float}

\definecolor{TUMpurple}{RGB}{105,008,090}
\definecolor{TUMviolet}{RGB}{015,027,095}
\definecolor{TUMpantone283}{RGB}{152,198,234} 
\definecolor{TUMpantone542}{RGB}{100,160,200} 
\definecolor{TUMpantone300}{RGB}{000,101,189} 
\definecolor{TUMpantone285}{RGB}{000,115,207} 
\definecolor{TUMpantone301}{RGB}{000,082,147} 
\definecolor{TUMpantone540}{RGB}{000,051,089} 

\definecolor{TUMcyan}{RGB}{000,119,138}    

\definecolor{TUMgreen}{RGB}{000,124,48}
\definecolor{TUMlightgreen}{RGB}{103,154,29}

\definecolor{TUMyellow}{RGB}{255,220,000}
\definecolor{TUMgold}{RGB}{249,186,000}
\definecolor{TUMorange}{RGB}{227,114,034}
\definecolor{TUMorangered}{RGB}{214,076,019}
\definecolor{TUMred}{RGB}{196,007,027}
\definecolor{TUMdarkred}{RGB}{156,013,022}

\definecolor{ownGray}{HTML}{707070}%
\definecolor{ownOrange}{HTML}{EA801C}%
\definecolor{ownPurple}{HTML}{800074}%

\definecolor{ownRedDark}{HTML}{A00000}%
\definecolor{ownRedMed}{HTML}{C46666}%
\definecolor{ownRedLight}{HTML}{D8A6A6}%

\newsavebox{\sampler}
\savebox{\sampler}{%
    \begin{tikzpicture}[baseline]
        \centering
        \node[anchor=center] (samp1) at (0,0) {};
        \node[anchor=center] (samp2) at ($(samp1) + (0.5,0)$) {};
        \node[anchor=center] (samp3) at ($(samp2) + (0.6,0.5)$) {};
        \node[anchor=center] (samp4) at ($(samp3) + (0.15,-0.5)$) {};
        \node[anchor=center] (samp5) at ($(samp4) + (0.5,0)$) {};
        
        \draw[] (samp1.west) -- (samp2);
        \draw[] (samp2.west) -- (samp3);
        \draw[] (samp4.west) -- (samp5);
        
        \node[] (1) at ($(samp2) + (-0.3,0.25)$) {};
        \node[] (2) at ($(samp4) - (0.2,0.1)$) {};
        \draw [-latex] (1) to [out=20] (2);
    \end{tikzpicture}%
}

\makeatletter

\newcommand*\wthelper[2]{%
        \hbox{\dimen@\accentfontxheight#1%
                \accentfontxheight#11.15\dimen@
                $\m@th#1\widetilde{#2}$%
                \accentfontxheight#1\dimen@
        }%
}

\newcommand*\accentfontxheight[1]{%
        \fontdimen5\ifx#1\displaystyle
                \textfont
        \else\ifx#1\textstyle
                \textfont
        \else\ifx#1\scriptstyle
                \scriptfont
        \else
                \scriptscriptfont
        \fi\fi\fi3
}
\makeatother

\newenvironment{customlegend}[1][]{%
    \begingroup
    \csname pgfplots@init@cleared@structures\endcsname
    \pgfplotsset{#1}%
}{%
    \csname pgfplots@createlegend\endcsname
    \endgroup
}%
\def\addlegendimage{\csname pgfplots@addlegendimage\endcsname}

\NewDocumentCommand{\addlegendimageintext}{o m}{%
    \tikz {
        \begin{customlegend}[
            legend entries={\empty},
            legend style={
                draw=none,
                line width=1.0pt,
                inner sep=0pt,
                column sep=0pt,
                mark size=2pt,
                mark options={solid},
                nodes={inner sep=0pt}}]
        \IfValueTF{#1}{\addlegendimage{#1,#2}}{\addlegendimage{#2}}
        \end{customlegend}
    }%
}

\DeclareMathOperator{\E}{\mathbb{E}}

\begin{document}

\title{Low-Cost Phase Precoding for Short-Reach\\ Fiber Links with Direct Detection
{}
}

\author{Ulrike Höfler, Daniel Plabst and Norbert Hanik,~\IEEEmembership{Senior Member,~IEEE}
\thanks{
U. Höfler, D. Plabst and N. Hanik are with the Institute for Communications Engineering, Technical University of Munich, 80333
Munich, Germany\\(e-mail: \{ulrike.hoefler, daniel.plabst, norbert.hanik\}@tum.de).
}}

\maketitle

\newcommand{\dan}[1]{\textit{\textcolor{blue}{[#1]}}}
\newcommand{\ulrike}[1]{\textit{\textcolor{red}{[#1]}}}

\begin{abstract}
Low-cost analog phase precoding is used to compensate chromatic dispersion (CD) in fibers with intensity modulation and direct detection (IM/DD). In contrast to conventional precoding with an in-phase and quadrature (IQ) Mach-Zehnder modulator (MZM), only a single additional phase modulator (PM) is required at the transmitter. Depending on the CD, the PM generates a periodic phase modulation that is modelled by a Fourier series and optimized via a mean squared error (MSE) cost criterion. Numerical results compare achievable information rates (AIRs) for 4- and 6-PAM. With the additional PM, energy gains of up to 3\,dB are achieved for moderate fiber lengths. 
\end{abstract}
\allowdisplaybreaks
\begin{IEEEkeywords}
intensity modulation/direct detection, short-reach fiber-optic, phase precoding, chromatic dispersion
\end{IEEEkeywords}

\section{Introduction}
Short-reach fiber-optic links with several to tens of kilometers are commonly used to connect data centers and metropolitan areas\cite{zhang2018performance,chagnon2019optical}.
To reduce hardware cost, these systems often use a direct detection (DD) receiver, which has a single photo-diode (PD) performing the optical to electrical conversion~\cite{ip2008coherent}. Short-reach fiber links with DD are mainly impaired by intersymbol interference (ISI) due to chromatic dispersion (CD), and the nonlinear squaring due to the PD~\cite{chagnon2019optical}. 

Dispersion can be compensated electrically or optically. Optical compensation techniques, such as dispersion compensating fibers (DCFs) add significant link loss \cite{gruner2005dispersion,morsy201850} and additional hardware cost. In the electric domain, one may compensate dispersion via receiver equalization or transmitter precoding. We briefly discuss these methods in the following subsections \ref{ssec:A} and \ref{ssec:B}.

\subsection{Equalization}\label{ssec:A}
Electronic linear equalizers~\cite{plabst2020wiener} cannot cancel the noninear squaring operation of the PD and lose rate at high-SNR. Nonlinear equalizers such as decision feedback equalizers (DFEs), Volterra filters, forward-backward algorithm (FBA) equalizers or neural network (NN) equalizers~\cite{plabst2024neural} show good performance in general. However, DFEs lead to error propagation and add complexity via an additional feedback filter~\cite{wettlin2020dsp}. Volterra filters and FBA-based approaches are generally complex. NN-based equalization~\cite{plabst2024neural} shows good rate and complexity improvements, but simpler compensation techniques may be preferred for low-cost and low-latency short-reach links with DD.

\subsection{Precoding}\label{ssec:B}
One may precode the transmit signal to compensate the complex-valued CD. The papers~\cite{killey2005electronic2,killey2006electronic,fonseca2005transmission,goeger2006modulation} use coherent precoding with two Mach-Zehnder modulators (MZM) and compensate the CD electrically via linear transmitter zero-forcing. 
To reduce hardware cost,~\cite{killey2005electronic1} performs CD compensation with a single dual-drive MZM. The approach is interference-limited, as the dual-drive configuration does not allow independent precoding of the signal amplitude and phase, and thus cannot fully precompensate the CD.
The authors of~\cite{erro1999novel} consider hybrid optical and electrical CD compensation for Gaussian transmit pulses. A fiber Bragg grating (FBG) is used to optically precompensate a major part of the CD. Subsequently, an electrically  tuneable phase modulator (PM) is configured to remove the remaining dispersion. 

Other works use directly modulated lasers (DMLs) or electro-absorption modulated lasers (EMLs) for intensity modulation~\cite{mitomi1994chirping,jamro2005chirp,jamro2007optimising,krehlik2007directly,del2008directly,sato2005chirp}.
In IM systems, DMLs introduce a time-dependent phase to the bandpass signal, which is proportional to both the output power and the rate of change in output power. In contrast, EMLs introduce phase modulation that depends solely on the rate of change in output power~\cite[Sec.~2.1]{zhang2018performance}.
Notably, this generated time-varying phase can be exploited for partial precompensation of CD. For low data rates around $\SI{10}{\giga\bit/\second}$, the authors of~\cite{henmi1994prechirp} use the time-dependent phase introduced by the DML for symbol-wise phase precoding in combination with an electro-absorption modulator (EAM) for IM. However, the performance of DMLs is significantly affected by intrinsic characteristics such as modulation efficiency and residual intensity modulation, limiting their suitability for this approach at higher data rates~\cite{preuschoff2022wideband,wyrwas2009dynamic,imai1990measurement,kobayashi1982direct,sanaee2017enhanced}.
The paper~\cite{jamro2005chirp} uses an EML with an optimized operating point to mitigate the effects of CD.
Additionally, the authors of~\cite{walker2005960} jointly optimize the DML and EAM for modulation and CD compensation. At high data rates, however, additional receiver equalization is required to compensate the interaction between the nonlinear modulators, the larger CD and the PD~\cite{zhang2018performance}. EMLs, for instance, exhibit stronger nonlinear behavior at higher IM levels because the resulting time-dependent phase varies with different modulation voltage levels, leading to further nonlinear distortions when interacting with the CD.
Similarly, for DMLs, the introduced phase modulation creates a power-dependent carrier frequency shift, which is converted into additional nonlinear distortions by CD~\cite[Sec.~2.2]{zhang2018performance }.
\clearpage
\subsection{Contributions and Organization}
This paper considers low-cost analog phase precoding for CD compensation. The transmitter uses a single MZM for amplitude modulation, followed by a PM that performs symbol-wise periodic phase precoding. The phase precoding depends on the total CD, transmit filter, the statistics of the transmit constellation and the receive filter. The periodic control signal to the PM may be generated via low-cost analog circuits, e.g., oscillators, frequency doublers and operational amplifiers. 

In contrast to~\cite{erro1999novel}, we model the periodic phase by a Fourier series and obtain a closed-form expression of the mean-squared error (MSE) between transmitted and received symbols. The coefficients of the PM are optimized for frequency domain raised cosine (FD-RC) and time domain raised cosine pulses (TD-RC) as well as 4- and 6-PAM.

The paper~\cite{xu2022flat} shows that a MZM and a PM can be integrated in a single compact and low-loss modulator package. Thus, the presented precompensation scheme may be a good candidate for practical systems.

\subsection{Notation}
Matrices and vectors are written using bold letters, e.g., $\mathbf{x}$. The transpose, complex conjugate, and complex conjugate transpose (Hermitian) of $\mathbf{x}$ are $\mathbf{x}^\mathsf{T}$, $\mathbf{x}^\mathsf{*}$ and $\mathbf{x}^\mathsf{H}$, 
respectively. The Hadamard product between vectors $\mathbf{x}$ and $\mathbf{y}$ is written as  $\mathbf{x} \circ \mathbf{y}$. We use $|\mathbf{x}|$ to denote the element-wise absolute value of $\mathbf{x}$ and $\mathbf{x}^{\circ 2}$ denotes element-wise squaring. 

By $\mathbf{A} = \mathrm{diag}(\mathbf{a})$ we denote a diagonal matrix that has $\mathbf{a}$ on its diagonal, while $ \mathbf{a} = \mathrm{diag}(\mathbf{A})$ arranges the diagonal entries of $\mathbf{A}$ in the column vector $\mathbf{a}$.
The $N\times N$-dimensional identity matrix is denoted as $\mathbf{I}_N$ and the $N$-dimensional all-ones column vector is $\mathbf{1}_N$. 
By $\mathbf{e}_i$ we denote the unit column vector that contains zeros everywhere, expect at position $i+1$, where it contains a one.

We define the sinc function as $\mathrm{sinc}(t) = \sin(\pi t)/ (\pi t)$. The Fourier transform of the signal $x(t)$ is defined as $\mathcal{F}\{x(t)\}(f) = \int_{-\infty}^\infty x(t) e^{-\mathrm{j} 2\pi f t} \mathrm{d} t$ and convolution between two signals $x(t)$ and $y(t)$ is written as $(x * y)(t)$.
\section{System Model}
\subsection{Continuous-Time Model}
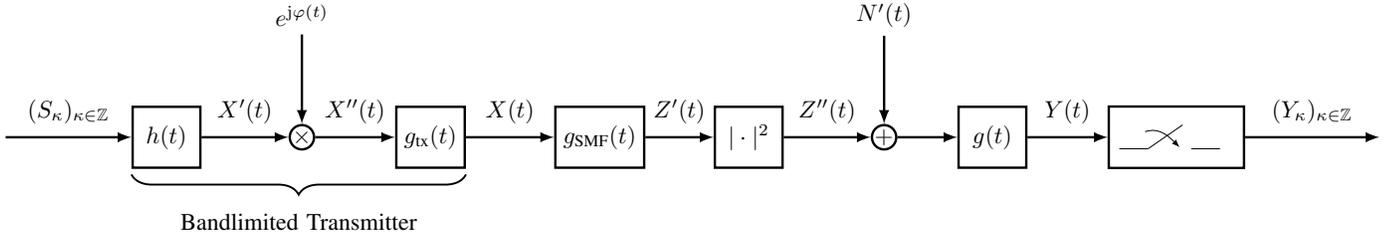
\begin{figure*}
    \centering
    \scalebox{0.9}{
    \begin{tikzpicture}
        \node at (0,0) (in) {};

        \node[draw,minimum width=1cm,minimum height=.9cm,line width=1] (h) at (2.5,0) {$h(t)$};
        \draw[-latex,thick,line width=1] (in) -- node[midway,above=0.2em ]{$(S_\kappa)_{\kappa \in \mathbb{Z}}$}(h.west);

        \node[circle, draw,minimum size=0.1pt,inner sep=0pt,line width=1] (m) at ($(h) + (2,0)$) {$\times$};
        \draw[-latex,thick,line width=1] (h) --  node[midway,above=0.2em ]{$X'(t)$}(m.west);

        \node at ($(m)+(0,1.8)$) (pha) {$e^{\mathrm{j}\varphi(t)}$};
        \draw[-latex,thick,line width=1] (pha) -- (m);

        \node[draw,minimum width=1cm,minimum height=.9cm,line width=1] (txfilt) at ($(m) + (1.9,0)$) {$g_\text{tx}(t)$};
        
        \draw [thick,decorate,decoration={brace,amplitude=3mm,mirror,raise=.5mm,}] ($(h.west) + (0,-0.5)$) -- node[midway,below,yshift=-0.5cm]{Bandlimited Transmitter} ($(txfilt.east) + (0,-0.5)$);

        \node[draw,minimum width=1cm,minimum height=.9cm,line width=1] (ch) at ($(txfilt) + (2.5,0)$) {$g_\text{SMF}(t)$};
        
        \draw[-latex,thick,line width=1] (m.east) -- node[midway,above=0.2em ]{$X''(t)$}(txfilt.west);
        
        \node[draw,minimum width=1cm,minimum height=.9cm,line width=1] (sld) at ($(ch) + (2.2,0)$) {$|\cdot| ^2$};
        
        \draw[-latex,thick,line width=1] (txfilt) -- node[midway,above=0.2em ]{$X(t)$}(ch.west);
        \draw[-latex,thick,line width=1] (ch) -- node[midway,above=0.2em ]{$Z'(t)$}(sld);
        
        \node[circle, draw,minimum size=0.1pt,inner sep=0pt,line width=1] (p) at ($ (sld) + (2,0) $) {$+$};
        \draw[-latex,thick,line width=1] (sld) -- node[midway,above=0.2em ]{$Z''(t)$}(p.west);
        \node at ($(p)+(0,1.8)$) (n) {$N'(t)$};
        \draw[-latex,thick,line width=1] (n) -- (p);
        
        \node[draw,minimum width=1cm,minimum height=.9cm,line width=1] (g) at ($(p) + (1.6,0)$) {$g(t)$};
        \draw[-latex,thick,line width=1] (p.east) -- node[midway,above=0.2em ]{}(g.west);
        
        \node[anchor=center] (samp1) at ($(g.east) + (1,0)$) {};
        
        \node[draw,minimum width=1cm,minimum height=.9cm,line width=1,scale=0.85, inner xsep=5pt] (samp) at ($ (samp1) +(1.2,0)$) {\usebox{\sampler}};
        \draw[-latex,thick,line width=1] (g.east) -- node[midway,above=0.2em ]{$Y(t)$}(samp.west);
        
        \draw[-latex,thick,line width=1] (samp.east) -- node[midway,above=0.2em ]{$(Y_\kappa)_{\kappa \in \mathbb{Z}}$}($(samp)+(3,0)$);
        
    \end{tikzpicture}
    }
    \caption{Baseband model of the considered short reach optical communication system.}
    \label{fig:sysModel}
\end{figure*}
Figure~\ref{fig:sysModel} shows the considered short-reach fiber-optic system. The transmit symbols $(S_\kappa)_{\kappa \in \mathbb{Z}}$ are uniformly, independently and identically distributed $M$-PAM symbols with alphabet $\mathcal{A}_M = \{a_0, \ldots, a_{M-1}\}$. The digital-to-analog converter (DAC) performs pulse shaping with $h(t)$ and generates
\begin{align}
    X'(t) = \sum\nolimits_{\kappa\in\mathbb{Z}} S_{\kappa} \, h(t-\kappa T_\text{s})
\end{align}
with symbol rate $B = 1/T_\text{s}$. We focus on pulses $h(t)$ that are either FD-RC or TD-RC pulses with a roll-off factor of $\alpha_\mathrm{ps}$.
For instance, choosing FD-RC with $\alpha_\mathrm{ps} = 0$ gives the sinc pulse 
\begin{align}
    h(t) = \mathrm{sinc}\left(t/T_\text{s}\right)
\end{align}
with single-sided bandwidth $B/2$ ~\cite[Eq.~(6.17)]{gallager2008principles},
whereas choosing TD-RC\footnote{A TD-RC pulse is obtained by swapping $(t,T_\text{s})$ and $(f,B)$ in ~\cite[Eq.~(6.17)]{gallager2008principles}} with $\alpha_\mathrm{ps} = 0$ corresponds to a rectangular pulse 
\begin{align}
h(t) = 
\begin{cases}
1 \;,&\quad |t| \leq T_\text{s}/2 \\ 
0 \;,& \quad \text{otherwise}.
\end{cases}
\end{align} 

The pulse-shaped signal is precoded with a PM 
\begin{align}
   X''(t) =  X'(t) \, \mathrm{e}^{\mathrm{j}\varphi(t)}
   .
   \label{eq:general_phase_prec}
\end{align}
The choice of $\varphi(t)$ is discussed in the next subsection. 
We assume the transmitter to be band-limited, as this is useful for wavelength division multiplexing applications. The band limitation is modeled by a low-pass filter $G_\mathrm{tx}(f)$ and for simplicity we choose the brickwall filter 
\begin{align}
    G_\mathrm{tx}(f) 
    = \left\{
    \begin{aligned}
        & 1, && |f| \leq B_\text{tx}/2\\
        & 0, && \text{otherwise}. \\
    \end{aligned}
    \right.
    \label{eq:bandlim}
\end{align}

The low-pass filtered transmit signal $X(t)$ passes through the linear fiber channel~\eqref{eq:cd_response} with the baseband transfer function: 
\begin{align}
    H_\mathrm{SMF}(f) = \mathrm{e}^{-\mathrm{j}\frac{\beta_2}{2} (2\pi f)^2 L}\,\mathrm{e}^{-\frac{\alpha_\mathrm{SMF}}{2}L},
    \label{eq:cd_response}
\end{align}
where $\beta_2$ is the chromatic dispersion coefficient, $\alpha_\mathrm{SMF}$ the attenuation coefficient and $L$ the fiber length.

At the receiver, a PD performs an optical-to-electrical conversion and generates $Z''(t) = |Z'(t)|^2$. We consider the PD at the thermal noise limit, where stationary real Gaussian noise $N'(t)$ with a two-sided power spectral density (PSD) of $\Phi_{N'}(f) = N_0/2$ is added; see Fig.~\ref{fig:sysModel}.

The receive filter limits the noise power and is modelled as a unit-gain brickwall filter $g(t)$ with two-sided bandwidth $B_\text{rx}$. The noise $N(t) = (g * N')(t)$ is colored zero-mean Gaussian with average power $\sigma_N^2 =B_\text{rx}\cdot N_0/2 $ and PSD
\begin{align}
    \Phi_{N}(f) = |G(f)|^2 \, \Phi_{N'}(f)  = \left\{
    \begin{aligned}
        & \frac{N_0}{2}, && |f| \leq B_\mathrm{rx}/2\\
        &0, &&\text{otherwise.} 
    \end{aligned}
    \right.
\end{align}

\subsection{Phase Modulator}
\begin{figure}[H]
\centering
    \begin{tikzpicture}
    \begin{axis}[
        height=6cm,
        width=0.48*\textwidth,
        xlabel=$t/T_s$,
        xmin=-3.1, xmax=2.5,
        ymin=-0.5, ymax=2.5,
        axis lines=center,
        axis on top=true,
        domain=-5:5,
        samples=200,
        legend pos=outer north east,
        xtick={-3,-2,-1,0,1,2},
        xticklabels={},
        x tick style={thick},
        ytick=\empty,
        axis y line=none,
        extra x ticks={0},
        extra x tick labels={},
        legend cell align={left}, 
        legend style={at={(0.6,0.8)},draw=none, font=\footnotesize},
        every axis x label/.style={
        at={(ticklabel* cs:1.1)},
        anchor=east,
        xshift=10pt,
      },
    ]

    \addplot[forget plot,black,dashed, ,domain=-3:2]
    {(sin(deg(pi*(x+1))) / (pi*(x+1))) * (cos(deg(pi*(x+1))) / (1 - (2*(x+1))^2))};
    
    \addplot[forget plot,black, ,domain=-3:2] {(sin(deg(pi*(x))) / (pi*(x))) * (cos(deg(pi*(x))) / (1 - (2*(x))^2))};
    
    \addplot[red, thick,domain=-3:2] {(sin(deg(pi*(x+1))) / (pi*(x+1))) * (cos(deg(pi*(x+1))) / (1 - (2*(x+1))^2))+(sin(deg(pi*(x))) / (pi*(x))) * (cos(deg(pi*(x))) / (1 - (2*(x))^2))};
    \addplot[blue,line width=1.0pt,domain=-3:2] {10}; 
    
    \addplot[black, ycomb,densely dotted] coordinates {(-2,-0.8) (-2,1)};
    \addplot[black, ycomb,densely dotted,] coordinates {(-1,-0.8) (-1,1)};
    \addplot[black, ycomb,densely dotted] coordinates {(0,-0.8) (0,1)};
    \addplot[black, ycomb,densely dotted] coordinates {(1,-0.8) (1,1)};
    
    \addlegendentry{Baseband Signal $X'(t)$};
    \addlegendentry{Phase Precoding $\varphi(t)$};
    \end{axis}

     \begin{axis}[
        yshift = -2cm,
        height=5cm,
        width=0.48*\textwidth,
        xlabel=$t/T_s$,
        xmin=-3.1, xmax=2.5,
        ymin=-0.5, ymax=1.5,
        axis lines=center,
        axis on top=true,
        domain=-5:5,
        samples=200,
        legend pos=outer north east,
        xtick={-3,-2,-1,0,1,2},
        xticklabels={},
        x tick style={thick},
        ytick=\empty,
        axis y line=none,
        extra x ticks={-1,0},
        extra x tick labels={$\kappa$,$\kappa+1$},
        every x tick label/.append style = {font=\footnotesize},
        every axis x label/.style={
        at={(ticklabel* cs:1.1)},
        anchor=east,
        xshift=10pt,
      },
    ]
    
    \addplot[blue,line width=1.0pt,domain=-3:2] {0.3*(cos(deg(2*pi*x)))-0.2*(cos(deg(2*pi*x*2)))-0.05*(cos(deg(2*pi*x*3)))};
   
    \addplot[black, ycomb,densely dotted] coordinates {(-2,-0.0) (-2,1)};
    \addplot[black, ycomb,densely dotted,] coordinates {(-1,-0.0) (-1,1)};
    \addplot[black, ycomb,densely dotted] coordinates {(0,-0.0) (0,1)};
    \addplot[black, ycomb,densely dotted] coordinates {(1,-0.0) (1,1)};

    \end{axis}
\end{tikzpicture}
    \caption{Example of a baseband signal $X'(t)$ with FD-RC $\alpha_\mathrm{ps}=1$ and symbols $(+1,+1)$ at times $(\kappa, \kappa+1)$. The analog phase precoding $\varphi(t)$ is $T_\text{s}$-periodic, exemplarily with $\eta_i \neq 0,\, i \in \{1,2\}$.}
    \label{fig:generalConcept}
\end{figure}
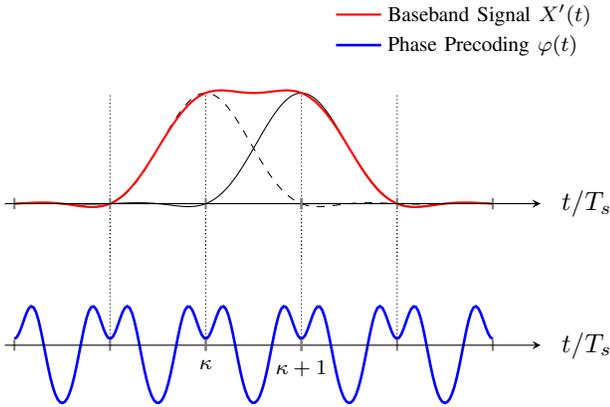

We choose an even, $T_\text{s}$-periodic phase $\varphi(t)$, expressed via the Fourier series\footnote{We dropped the constant $\eta_0$, as the PD output $Z''(t)$ is invariant to constant phase rotations. We also neglect the $\sin(\cdot)$ terms in the Fourier series, as optimizing the associated weights corresponding to section \ref{sec:opti} results in zero weighting.}:   
\begin{align}\label{eq:generalPhase}
    \varphi(t) = \sum\nolimits_{w=1}^{W}\eta_w\,\mathrm{cos}\left(2\pi wtB\right) 
\end{align}
with coefficients $\eta_w$ $\in \mathbb{R}$ for $w \in \{1,\ldots,W\}$. The $W$ coefficients adjust the phase modulation; see Fig.~\ref{fig:generalConcept}.
Phase precoding, being inherently nonlinear, causes spectral expansion. The Fourier transform of~\eqref{eq:general_phase_prec} is 
\begin{align}\label{eq:chirpedSignal}
    X''(f) = \Big(X'  * \mathcal{F}\Big\{\exp{\big(\mathrm{j} \varphi(t)\big)}\Big\}\Big)\big(f\big). 
\end{align}
We use~\eqref{eq:generalPhase} in~\eqref{eq:chirpedSignal} to compute the frequency response of the phase modulation
\begin{equation}
\begin{aligned}
    \mathcal{F}\{e^{\mathrm{j} \varphi(t)}\}(f) &= 
     \mathcal{F} \big\{ \prod\nolimits_{w=1}^W   e^{\mathrm{j} \,\eta_w \cos(2 \pi w t B) }  \big\}(f)\\
    &=  (\underbrace{\Omega_1 *  \cdots *  \Omega_w * \cdots * \Omega_W}_{W \times})(f),
\end{aligned}
\end{equation}
where the last step follows from the Jacobi-Anger expansion for complex exponentials 
\begin{equation}
\begin{aligned}
    e^{\mathrm{j} \eta_w \cos(2 \pi w t B)} &= \sum\nolimits_{u\in\mathbb{Z}} \mathrm{j}^u J_u(\eta_w) e^{\mathrm{j} 2 \pi  u   w t B} 
    \label{eq:jangerexp}
\end{aligned}
\end{equation}
and the Fourier transform of~\eqref{eq:jangerexp}:
\begin{equation}
\begin{aligned}   
    \Omega_w(f) &=  \sum\nolimits_{u\in\mathbb{Z}} \mathrm{j}^{u} J_{u}(\eta_w)\,\delta(f - w u B),
    \label{eq:fourier_omegan}
\end{aligned}
\end{equation}
where the Bessel function of the first kind  of $u$-th order is denoted by $J_u(\eta_w)$. 
\vspace{-1.2cm}
As an example, consider the phase precoding~\eqref{eq:generalPhase} with one non-zero Fourier coefficient $(\eta_1)$. The spectrum
\begin{align}
   X''(f) = \sum\nolimits_{u\in\mathbb{Z}}\mathrm{j}^{u} J_{u}(\eta_1)\, X'(f - u B)
    \label{eq:pm_spectrum}
\end{align}
has unbounded bandwidth, as weighted spectral replica of the baseband signal $X'(f)$ appear centered at $uB$, see~\eqref{eq:fourier_omegan}.
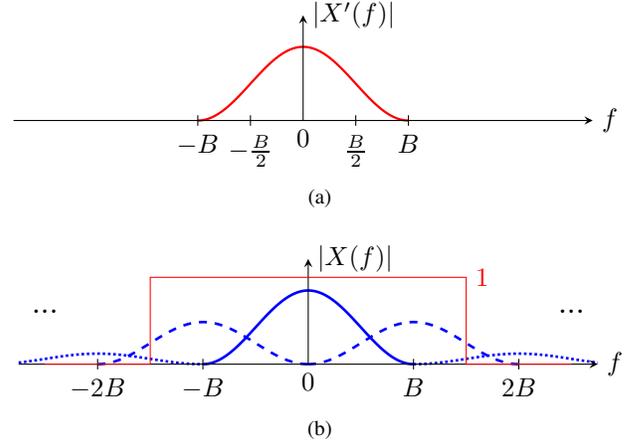
\begin{figure}[!ht]
    \centering
    \subfloat[]{
        \tikzset{mylines/.style={line width=0.3mm,}}

\begin{tikzpicture}[scale=0.7]

    \draw[red, mylines, domain=-2:2, variable=\x, red, samples=200, smooth] plot ({\x}, {0.7*(1+cos(deg(pi/2*\x))});
    
    \draw[-stealth] (-5.5,0) -- (5.5,0) node[right] {$\scalebox{1}{$f$}$};
    \draw[-stealth] (0,0) node[below] {$\scalebox{1}{$0$}$} -- (0,2) node[ right] {$\scalebox{1}{$|X'(f)|$}$};

    \draw (-1,0.1) -- (-1,-0.1) node[below] {$\scalebox{1}{$-\frac{B}{2}$}$};
    \draw (1,0.1) -- (1,-0.1) node[below] {$\scalebox{1}{$\frac{B}{2}$}$};
    
    \draw (-2,0.1) -- (-2,-0.1) node[below] {$\scalebox{1}{$-B$}$};
    \draw (2,0.1) -- (2,-0.1) node[below] {$\scalebox{1}{$B$}$};


    
    
        


\end{tikzpicture}
        \label{fig:spectrumA}
    }
    \hfill
    \subfloat[]{
        \newcommand\freq{1.1}

\tikzset{mylines/.style={line width=1pt,}}

\begin{tikzpicture}[scale=0.7]
    \coordinate (pic4) at (0,-3.5);
    \begin{scope}[shift={(pic4)}]

        \draw[blue, mylines, domain=-2:2, variable=\x, blue, samples=200, smooth] plot ({\x}, {0.7*(1+cos(deg(pi/2*\x))});

        \draw[red, mylines, domain=0:4, variable=\x, blue, dashed, samples=200, smooth,line width=1pt] plot ({\x}, {0.4*(1+cos(deg(pi/2*(\x-2)))});
        \draw[red, mylines, domain=-4:0, variable=\x, blue, dashed, samples=200, smooth] plot ({\x}, {0.4*(1+cos(deg(pi/2*(\x+2)))});

        \draw[blue,densely dotted,mylines, domain=2:5.5, variable=\x, samples=200, smooth] plot ({\x}, {0.1*(1+cos(deg(pi/2*(\x-4)))});
        \draw[blue,densely dotted, mylines, domain=-5.5:-2, variable=\x,  samples=200, smooth] plot ({\x}, {0.1*(1+cos(deg(pi/2*(\x+4)))});
        
        \draw[-stealth] (-5.5,0) -- (5.5,0) node[right] {$\scalebox{1}{$f$}$};
        \draw[-stealth] (0,0) node[below] {$\scalebox{1}{$0$}$} -- (0,2) node[ right] {$\scalebox{1}{$|X(f)|$}$};
    
        \draw (-2,0.1) -- (-2,-0.1) node[below] {$\scalebox{1}{$-B$}$};
        \draw (2,0.1) -- (2,-0.1) node[below] {$\scalebox{1}{$B$}$};
    
        \draw (-4,0.1) -- (-4,-0.1) node[below] {$\scalebox{1}{$-2B$}$};
        \draw (4,0.1) -- (4,-0.1) node[below] {$\scalebox{1}{$2B$}$};
        
        \node at (-5,1) {{$\scalebox{1.2}{$...$}$}};
        \node at (5,1) {{$\scalebox{1.2}{$...$}$}};

        \draw[red, solid] (-5,0) -- (-3,0) -- (-3,1.65) -- (3,1.65) node[right]{$1$} -- (3,0) -- (5,0); 
        
    \end{scope}

\end{tikzpicture}
        \label{fig:spectrumB}
    }
    \caption{Spectrum of a FD-RC $\alpha_\mathrm{ps} = 1$: (a) before phase precoding and (b) after phase precoding. The frequency response of a transmit filter with bandwidth $B_\text{tx} = 3 B$ is shown in red.}
    \label{fig:spectrum}
\end{figure}

Applying the filter\eqref{eq:bandlim} to~\eqref{eq:chirpedSignal} results in 
\begin{align}
   X(f) = G_\text{tx}(f)\sum_{u^W} \bigg(\prod_{i=1}^W \mathrm{j}^{u_i} J_{u_i}(\eta_i)\bigg)  X'\bigg(f - B \sum_{i=1}^W i  u_i\bigg),
    \label{eq:filtered_pm_spectrum}
\end{align}
where $u^W := (u_1, \ldots, u_W)$ and $u_i \in \mathbb{Z}$ for $i=\{1,\cdots,W\}$.
We set the bandwidth of~\eqref{eq:bandlim} to $B_\text{tx} := B\left( 2F - 1 \right)$ and $F \in  \mathbb{N}^+$. 
For instance, for a transmit signal spectrum $X'(f)$ that is bandlimited to $|f|\leq B$, setting $F=2$ will produce a PM output $X(f)$ depending on the four weighted spectral copies of $X'(f)$ centered at $f \in \lbrace \pm B, \pm 2\,B \rbrace$; see Fig.~\ref{fig:spectrum}.
\subsection{Discrete-Time Model}
To obtain a low-complexity receiver, we sample $Y(t)$ at symbol rate\footnote{If the filtered signal $Y(t)$ occupies a larger bandwidth than $B$, sampling at symbol rate may not result in sufficient statics and may thus cause an achievable rate loss; see~\cite[Fig.~7]{plabst2022achievable}. One can find a trade-off between rate loss and receiver complexity.}, i.e., we set $t =  \kappa T_\text{s}$, $\kappa \in \mathbb{Z}$. 
\begin{align}
    Y_\kappa = Z_\kappa + N_\kappa,
    \label{eq:yk_generel}
\end{align}
where
\begin{align}
    Z_\kappa &= 
    \left[
    g(t) * |Z'(t)|^2
    \right]_{t = \kappa T_\text{s}} 
    \label{eq:zk}
    \\
    N_\kappa &= 
    \left[
    g(t) * N'(t)
    \right]_{t = \kappa T_\text{s}}
    .
\end{align}
In general, the noise $N_\kappa$ is a real-valued discrete-time colored Gaussian process with ACF $\varphi_\text{NN}[\kappa] = \sigma_N^2 \cdot g(\kappa T_\text{s})$.
%
%
%
\subsection{Approximation via Simulation}
In order to obtain a discrete-time model for the channel inputs $(S_\kappa)_{\kappa \in \mathbb{Z}}$ and outputs $(Z_\kappa)_{\kappa\in  \mathbb{Z}}$, caution is required due to the bandwidth expansion of the nonlinear PM and PD.

The bandwidth of the PM output signal $X''(t)$ is in general unbounded, see~\eqref{eq:pm_spectrum}. The additional filter $g_\text{tx}(t)$ ensures that $X(t)$ is bandlimited to $\pm B_\text{tx}/2$; see Fig.~\ref{fig:spectrum}. Since the fiber is assumed to be linear, no spectral broadening takes place. 
Finally, the PD squaring operation doubles the bandwidth of $Z''(t)$ to $\pm B_\text{tx}$. We use oversampling to address these bandwidth expansions; see~\cite[Sec.~II.~D]{plabst2024neural}. 

Assume oversampling with factor $N_\text{os}$ and corresponding sampling period $T_\text{sim} = T_\text{s}/N_\text{os}$. The noise-free channel outputs are 
\begin{align}
    Z_\kappa = \sum_{\ell_1\in \mathbb{Z}} g_{\ell_1} \, \bigg| 
     \sum_{\ell_2 \in \mathbb{Z}} g_\text{SMF,$\ell_2$} \,X_{\kappa N_\mathrm{os} -\ell_1 - \ell_2}
    \bigg| ^2
    \label{eq:zk_sum}
\end{align}
where $g_\ell := g(\ell T_\text{sim})$,  $g_{\text{SMF},\ell} := g_\text{SMF}(\ell T_\text{sim})$, and $X_{\ell}:=X(\ell T_\text{sim})$. 
The expansion~\eqref{eq:filtered_pm_spectrum} gives insight into the spectrum of the PM output $X(t)$: As an example, consider $W=1$, a bandlimited DAC signal $X'(t)$ on $|f|\leq B/2$ and finite $B_\text{tx}$. In this case we may simplify~\eqref{eq:filtered_pm_spectrum} to
\begin{align}
   X(f) = \sum\nolimits_{u \in \mathbb{Z} \,:\, |u| \leq B_\text{tx}/(2B)} \,\mathrm{j}^{u} J_{u}(\eta_1)   X'\big(f - B u\big)
    \label{eq:filtered_pm_spectrum_2}
    .
\end{align}
One may generate discrete-time representations $X_\ell$ from the bandlimited~\eqref{eq:filtered_pm_spectrum_2}.
Unfortunately, for $W \geq 2$, the summation in~\eqref{eq:filtered_pm_spectrum} is over infinitely many terms, which is unpractical. Furthermore, for TD-RC pulses, the DAC signal $X'(t)$ is unbounded in frequency. Thus, for the general case, we approximate the bandlimited PM signal as  
\begin{align}
X_\ell \approx  
\sum_{\ell_1 \in\mathbb{Z}} 
g_{\text{tx},\ell_1} \mathrm{e}^{\mathrm{j}\varphi_{\ell-\ell_1}} \sum_{\ell_2 \in\mathbb{Z}}  h_{\ell_2} S'_{\ell -\ell_1 - \ell_2}
\label{eq:xk_approx}
\end{align}
with $g_{\text{tx},\ell} := g_\text{tx}(\ell T_\text{sim})$, $\varphi_\ell := \varphi(\ell T_\text{sim})$, $h_\ell := h(\ell T_\text{sim})$, and the $N_\text{os}$-fold upsampled symbol sequence $(S'_\ell)_{\ell \in \mathbb{Z}} := ((0,\ldots,0),S_\kappa)_{\kappa \in \mathbb{Z}}$. Eq.~\eqref{eq:xk_approx} is an approximation, as the PM signal $X''(t)$ has unbounded bandwidth in general\footnote{
For the values of $\eta_w$ considered in the numerical results, we saw through simulations that the weighting factors $J_u(\eta_w)$ decay quickly.}.

Note that~\eqref{eq:zk_sum} requires $N_\text{os} \geq 2 B_\text{tx}/B$ due to the squaring operation of the PD, while~\eqref{eq:xk_approx} may require a much larger oversampling factor. In the numerical results section, we choose $N_\text{os} \gg 2 B_\text{tx}/B$ based on a distortion  criterion, which ensures that both~\eqref{eq:zk_sum} and~\eqref{eq:xk_approx} are accurate.

\section{Optimization via MSE}\label{sec:opti}
Next, we formulate an MSE objective function for optimizing $\boldsymbol{\eta} := (\eta_i)_{i=1}^{W}$. We add a real-valued scaling parameter $\xi$ to the MSE cost function $\epsilon$ to take fiber attenuation into account. The scaling parameter can be interpreted as an automatic gain control. 
We define the MSE between a single transmitted and received symbol as
\begin{align}\label{eq:costfunction2}
    \begin{split}
       (\boldsymbol{\eta}^\star, \xi^\star) 
       =\underset{\boldsymbol{\eta},\xi}{\mathrm{arg\,min}}  \,\,\epsilon, \,\; \quad \epsilon := \E\left[(S_\kappa - \xi \cdot Y_\kappa )^2\right]
       .
   \end{split}
\end{align}
Next, we formulate $Y_\kappa$ using the matrix-vector representations of~\eqref{eq:zk_sum} and~\eqref{eq:xk_approx}. First, combine the transmit filter and fiber channel into $f(t) := (g_\mathrm{tx} * h_\mathrm{SMF})(t)$. 
Assume that the taps $h_{\ell}$, $f_\ell := f(\ell T_\text{sim})$ and $g_\ell$ are zero outside some interval and define the vectors 
\begin{align}
    \mathbf{f} &:=[f_{+(M_f-1)/2},\ldots f_0 \ldots,f_{-(M_f-1)/2}\big]^\mathsf{T} &&\in \mathbb{C}^{M_f}  
    \label{eq:vec_psi}
    \\
    \mathbf{g} &:= [g_{+(M_g-1)/2},\ldots,g_0, \ldots g_{-(M_g-1)/2}]^\mathsf{T} &&\in \mathbb{R}^{M_g}
    \label{eq:vec_g}
    \\
    \mathbf{h} &:= [h_{+(M_h-1)/2},\ldots,h_0 \ldots h_{-(M_h-1)/2}]^\mathsf{T} &&\in \mathbb{R}^{M_h}
    \label{eq:vec_h}
\end{align}
respectively, where $M_h$, $M_f$ and $M_g$ are odd integers. 
Define the Toeplitz matrix $\mathbf{F}\in\mathbb{C}^{M_g \times M_g + M_f - 1}$ that is constructed from right-shifted copies of $\mathbf{f}^\mathsf{T}$. In the same fashion, the matrix $\mathbf{H}'\in \mathbb{R}^{M_g + M_f - 1 \times Q}$, $Q:=  M_g + (M_f-1) + (M_h - 1)$ is constructed from right-shifted copies of $\mathbf{h}^\mathsf{T}$.

The channel outputs are 
\begin{align}\label{eq:yk}
    \begin{split}
        Y_\kappa
        &= \mathbf{g}^\mathsf{T}\,\big|\mathbf{F}\, 
        \left(
        \exp{(\mathrm{j} \boldsymbol{\varphi}')} \circ (\mathbf{H}' \mathbf{S}'_\kappa) 
        \right) \big|^{\circ 2}+ N_\kappa 
        ,
    \end{split}
\end{align}
with the upsampled sequence 
\begin{align}
   \mathbf{S}_\kappa' &= [ \underbrace{(S_{\kappa+i}, \mathbf{0}),}_{\text{Repeat $i = -B_1,\ldots1$}}S_\kappa, \underbrace{(\mathbf{0}, S_{\kappa+i})}_{\text{Repeat $i = 1,\ldots B_1$}}]^\mathsf{T}
\end{align}
where $\mathbf{0} := \mathbf{0}_{N_\text{os}-1}^\mathsf{T}$ and the periodically repeating phases 
\begin{align}\label{eq:varphiPrime}
     \boldsymbol{\varphi}' &=  [  
     \underbrace{(\varphi_0, \varphi_1 \ldots \varphi_{N_\text{os}-1})}_{\text{Repeat $B_2$ times}},\;
     \varphi_0,\;
     \underbrace{(\varphi_1, \ldots \varphi_{N_\text{os}-1},\varphi_0)}_{\text{Repeat $B_2$ times}}]^\mathsf{T} 
     .
\end{align}
and integers $B_1$, $B_2$ that satisfy\footnote{Alternatively, we pad these filters with zeros on both sides.} the filter lengths~\eqref{eq:vec_psi}-\eqref{eq:vec_h}. 
Finally, we express the phase samples $\varphi_\ell$, $\ell \in \{0, \ldots, N_\text{os}-1\}$ by the inner product $\varphi_\ell = \boldsymbol{\eta}^\mathsf{T} \, \boldsymbol{\gamma}_\ell$, $\boldsymbol{\eta} := [\eta_1,\ldots,\eta_W]^\mathsf{T}$ and $\boldsymbol{\gamma}_\ell := [\cos(2\pi \ell/N_\text{os}), \ldots, \cos(2\pi W \ell/N_\text{os})]^\mathsf{T}$.

In~\eqref{eq:yk}, we simplify the product $\mathbf{H}' \mathbf{S}'_\kappa = \mathbf{H} \mathbf{S}_\kappa$, i.e., we removed the columns in $\mathbf{H}'$ that are multiplied by a zero in $\mathbf{S}'_\kappa$ and defined $\mathbf{S}_\kappa := [S_{\kappa-B_1}, \ldots S_{\kappa + B_1}]^\mathsf{T}$, $\mathbf{H}\in \mathbb{R}^{(M_f+M_g-1)\times K}$ and $K = (2B_1 + 1)$.

We first optimize the scaling parameter $\xi$ by setting the derivative
\begin{align}
    \frac{\partial \epsilon}{\partial \xi} &= -2\,\E\left[S_\kappa\, Y_\kappa\right] +2\,\xi\,\E\left[Y_\kappa^2\right] 
\end{align}
equal to zero and solving for $\xi$, which results in the optimal
\begin{align}
\label{eq:xistar}
    \xi^\star = \E\left[S_\kappa\, Y_\kappa\right] \big/ \E\left[Y_\kappa^2\right]
    .
\end{align}
Insert~\eqref{eq:xistar} into $\epsilon$ to obtain  
\begin{align}\label{eq:costfunctionU}
    \begin{split}
        \epsilon &= \E[S_\kappa^2] -\E\left[S_\kappa\, Y_\kappa\right]^2 \big/ \E\left[Y_\kappa^2\right],
    \end{split}
\end{align}
where $Y_\kappa$ is a function of $\boldsymbol{\eta}$, and $\E[S_\kappa^2]$ is a constant with respect to $\boldsymbol{\eta}$ and can be neglected in the optimization. 

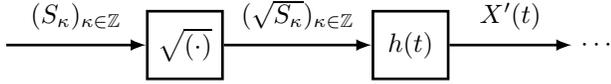
\begin{figure}
    \begin{tikzpicture}
    \node at (0,0) (in) {};

    \node[draw,minimum width=1cm,minimum height=.9cm,line width=1] (root) at (2.5,0) {$\sqrt{(\cdot)}$};
    \draw[-latex,thick,line width=1] (in) -- node[midway,above=0.2em ]{$(S_\kappa)_{\kappa \in \mathbb{Z}}$}(h.west);
    
    \node[draw,minimum width=1cm,minimum height=.9cm,line width=1] (h) at ($(h) + (3,0)$) {$h(t)$};
    \node (n) at ($(h) + (2.5,0)$) {$\cdots$};
    \draw[-latex,thick,line width=1] (h) --  node[midway,above=0.2em ]{$X'(t)$}(n.west);
    
    \draw[-latex,thick,line width=1] (root) -- node[midway,above=0.2em ]{$(\sqrt{S_\kappa})_{\kappa \in \mathbb{Z}}$}(h.west);
\end{tikzpicture}
    \caption{Transmitter with symbol-wise predistortion.}\label{fig:symbPred}
\end{figure}

When transmitting $M$-PAM, the elements in $\mathcal{A}_M$ are equidistant, while the noise-free channel outputs are not equidistantly spaced due to the PD squaring. Under AWGN, and at high-SNRs, equidistantly spaced channel outputs maximize the information rates. To create approximately equidistant channel outputs, we extend Fig.~\ref{fig:sysModel} by incorporating the square-root predistortion from~\cite{plabst2020wiener} and transmit $\sqrt{S_\kappa} \in \sqrt{\mathcal{A}_M} := \{\sqrt{a_0},...,\sqrt{a_{M-1}}\}$; see Fig.~\ref{fig:symbPred}. In principle, one can also optimize the constellation spacing for every SNR, which is called geometric shaping.

Define $\boldsymbol{\Psi}:=\mathbf{F}\, \mathrm{diag}(\exp{(\mathrm{j} \boldsymbol{\varphi}')}) \,\mathbf{H}$. Using the approximation~\cite[Sec.~IV.~A, B]{plabst2020wiener}, we obtain the expected values
\begin{align}
    \begin{split}
        \E[S_\kappa \, Y_\kappa] &=\mathbf{g}^\mathsf{T}\,\E\big[S_\kappa \cdot \big|\mathbf{F}\, \big[\exp{(\mathrm{j} \boldsymbol{\varphi}')} \circ (\mathbf{H}\, \sqrt{\mathbf{S}_\kappa})\big] \big|^{\circ 2}\big]\\
        &= \mathbf{g}^\mathsf{T}\,\E\big[S_\kappa  \cdot \big|\boldsymbol{\Psi}\, \sqrt{\mathbf{S}_\kappa}\big] \big|^{\circ 2}\big]\\
        &\approx  \mathbf{g}^\mathsf{T}\big(\mu_S^2\,\big|\mathbf{w}\big|^{\circ 2} + \frac{\sigma_S^2}{4}\, \mathbf{z}+\sigma_S^2\, \mathrm{Re}\big\{\mathbf{w}^\mathsf{*}\circ (\boldsymbol{\Psi}\, \mathbf{e}_{B_1})\big\}\big)
    \end{split}
\end{align}
and
\begin{align}\label{eq:EY2}
    \begin{split}
        \E[Y_\kappa^2] = \mathbf{g}^\mathsf{T} \underbrace{\E\big[\big|\boldsymbol{\Psi}  \sqrt{\mathbf{S}_\kappa}\big|^{\circ 2}\, \big|\boldsymbol{\Psi}  \sqrt{\mathbf{S}_\kappa}\big|^{\circ 2,\mathsf{T}}\big]}_{(a)} \, \mathbf{g} + \sigma^2_N,
    \end{split}
\end{align}
where the term $(a)$ is approximated by 
\begin{align}
    \begin{split}   
        &\mu_S^2 \, \big(|\mathbf{w}|^{\circ 2}\, |\mathbf{w}|^{\circ 2,\mathsf{T}}\big)\\
        &+ \frac{\sigma_S^2}{2}\,\big( \mathrm{Re}\big\{\mathrm{diag}(\mathbf{w}^\mathsf{*})\, \boldsymbol{\Psi}\, \boldsymbol{\Psi}^\mathsf{T}\, \mathrm{diag}(\mathbf{w}^\mathsf{*})\big\}\\
        & \qquad\quad+\mathrm{Re}\big\{\mathrm{diag}(\mathbf{w}^\mathsf{*})\, \boldsymbol{\Psi}\, \boldsymbol{\Psi}^\mathsf{H}\, \mathrm{diag}(\mathbf{w})\big\}\big)\\
        &+\frac{\sigma_S^2}{4}\,\big(\mathbf{z}\, \big|\mathbf{w}\big|^{\circ 2,\mathsf{T}} + \big|\mathbf{w}\big|^{\circ 2}\, \mathbf{z}^\mathsf{T}\big)\\
        &+\frac{\sigma_S^4}{16\,\mu_S^2}\, \big(\Big|\boldsymbol{\Psi}\, \boldsymbol{\Psi}^\mathsf{T}\big|^{\circ 2 } + \big|\boldsymbol{\Psi}\, \boldsymbol{\Psi}^\mathsf{H}\big|^{\circ 2 }+\mathbf{z}\, \mathbf{z}^\mathsf{T}\big)\\
        &+\frac{\mu_{4,S}-3\,\sigma_S^4}{16\,\mu_S^2}\, \big(\big|\boldsymbol{\Psi}\big|^{\circ 2}\, \big|\boldsymbol{\Psi}\big|^{\circ 2,\mathsf{T}}\big),
    \end{split}
\end{align}
see~\cite[Sec.~IV.~A, B]{plabst2020wiener}, and $\mathbf{w}:= \boldsymbol{\Psi}\, \mathbf{1}_K$, $\mathbf{z}:=\big|\boldsymbol{\Psi}\big|^{\circ 2}\, \mathbf{1}_K$. The mean, variance and the fourth order central moment of the random variable $S$ are denoted by  $\mu_S$,  $\sigma^2_S$ and $\mu_{4,S}=\E[(S-\mu_S)^4]$, respectively.
The gradient 
$\partial \epsilon / \partial \boldsymbol{\eta}$ of dimension $\mathbb{R}^{W}$ is provided in the appendix.

\section{Simulation Results}

\begin{figure*}
    \centering
    \subfloat[FD-RC with $L=\SI{1}{km}$.]{
        \scalebox{0.4}{\input{./pics/AIRoverSNR_RC_1.tex}}
    }
    \hspace{0.7cm}
    \subfloat[TD-RC with $L=\SI{1}{km}$.]{
        \scalebox{0.4}{\input{./pics/AIRoverSNR_RC_TD_1.tex}}
    }

    \vspace{-1.3cm}
    \subfloat[FD-RC with $L=\SI{2}{km}$.]{
        \scalebox{0.4}{
%
%
%
\begin{tikzpicture}
\pgfdeclarelayer{background}
\pgfdeclarelayer{foreground}
\pgfsetlayers{background,main,foreground}

\begin{axis}[%
width=7in,
height=3.8in,
at={(1.011in,0.642in)},
scale only axis,
xmin=-3.8,
xmax=15,
ymin=0,
ymax=log2(6),
axis background/.style={fill=white},
xtick={-3,-2,...,15},
ytick={0,0.2,...,2.6},
xticklabels={$-3$,,$-1$,,$1$,,$3$,,$5$,,$7$,,$9$,,$11$,,$13$,,$15$},
yticklabels={$0$,,$0.4$,,$0.8$,,$1.2$,,$1.6$,,$2$,,$2.4$,,$2.8$},
xticklabel style={scale=1.8},
yticklabel style={scale=1.8},
xlabel={SNR [dB]},
ylabel={AIR [bpcu]},
xlabel style={scale=2.5},
ylabel style={scale=2.5},
grid=both,
legend pos=south east,
legend style={nodes={scale=2, transform shape}}
]
\addplot [color=red, line width=1.0pt, mark=*,mark repeat=2, mark size = 3pt,mark phase = 2]
  table[row sep=crcr]{%
-15	-0.000739301337283763\\
-14.5	0.000152923689265598\\
-14	-0.000297938186064606\\
-13.5	0.000492441377399085\\
-13	0.00102735298200411\\
-12.5	0.000737229338697132\\
-12	0.000585057478331485\\
-11.5	0.0014945250930339\\
-11	0.00192750306850849\\
-10.5	0.00226658903098653\\
-10	0.00365748659096983\\
-9.5	0.00460721061505343\\
-9	0.00516552372877925\\
-8.5	0.00767966624157768\\
-8	0.0108071374082932\\
-7.5	0.0131737876810669\\
-7	0.0138569104371942\\
-6.5	0.0204473150060605\\
-6	0.0271445069294961\\
-5.5	0.0357543165131342\\
-5	0.0389349431254882\\
-4.5	0.0512224986798614\\
-4	0.0636013336584453\\
-3.5	0.0808965798900794\\
-3	0.0988167973274883\\
-2.5	0.124599823239497\\
-2	0.16012574246374\\
-1.5	0.193078849843954\\
-1	0.228280529995258\\
-0.499999999999998	0.272867450335322\\
2.89298239965986e-15	0.342107147789039\\
0.5	0.405659237337566\\
1	0.481011193212243\\
1.5	0.560958836368557\\
1.76091259055681	0.611336871965759\\
2	0.64514398779497\\
2.5	0.748776374508736\\
3	0.858959847983331\\
3.5	0.966517445763004\\
4	1.06620454339545\\
4.5	1.19583106556216\\
5	1.31450471497069\\
5.5	1.43985045634232\\
6	1.56832632918239\\
6.5	1.66827451819416\\
7	1.76726619407149\\
7.5	1.86007834137522\\
8	1.91850380149453\\
8.5	1.95486864986934\\
9	1.98038964910381\\
9.5	1.99322663953836\\
10	1.99839405337727\\
10.5	1.99955677719062\\
11	1.99998908625666\\
11.5	1.99999960694534\\
12	1.99999999973451\\
12.5	2\\
13	2\\
13.5	2\\
14	2\\
14.5	2\\
15	2\\
15.5	2\\
16	2\\
16.5	2\\
17	2\\
17.5	2\\
18	2\\
18.5	2\\
19	2\\
19.5	2\\
20	2\\
};

\addplot [color=red, line width=1.0pt, mark=*,mark repeat=2, mark size = 3pt,mark phase = 2,densely dashed]
  table[row sep=crcr]{%
-15	-0.00133262433980189\\
-14.5	-0.000806418239118016\\
-14	-0.000679488093378625\\
-13.5	-0.000510295948702444\\
-13	0.000219551309957581\\
-12.5	-0.000875026649984697\\
-12	-8.85541952446722e-05\\
-11.5	9.03006405510443e-05\\
-11	0.000878800257225532\\
-10.5	0.00187579061906658\\
-10	0.00278717354309525\\
-9.5	0.00318381230653812\\
-9	0.00445988573989542\\
-8.5	0.00526527302326703\\
-8	0.00798388960981367\\
-7.5	0.00941370190946304\\
-7	0.0120541207602742\\
-6.5	0.0156905443595458\\
-6	0.0209020395081462\\
-5.5	0.0262157467760048\\
-5	0.0323840638406207\\
-4.5	0.0416466029849518\\
-4	0.051190953275668\\
-3.5	0.0654349336557605\\
-3	0.0874180745933966\\
-2.5	0.104168337211308\\
-2	0.130901971053727\\
-1.5	0.167261117720797\\
-1	0.195673381333566\\
-0.499999999999998	0.238764024177281\\
9.64327466553287e-16	0.289725616220592\\
0.499999999999999	0.352805290009579\\
1	0.415011354920213\\
1.5	0.502620431991189\\
1.76091259055681	0.541134919304393\\
2	0.584556186244928\\
2.5	0.67308178946996\\
3	0.774477461556021\\
3.5	0.874462920395313\\
4	1.00165327648804\\
4.5	1.10513747702704\\
5	1.23852641347126\\
5.5	1.36209746887649\\
6	1.48705901075667\\
6.5	1.63128355579016\\
7	1.76340671021429\\
7.5	1.89955224117465\\
8	2.036538472896\\
8.5	2.16346106772882\\
9	2.28139324343847\\
9.5	2.39041358382019\\
10	2.46498845785384\\
10.5	2.51960953949612\\
11	2.55128161904572\\
11.5	2.57316295125466\\
12	2.58138936581203\\
12.5	2.5842275614492\\
13	2.58477023657619\\
13.5	2.58496044330979\\
14	2.5849624025012\\
14.5	2.58496250072092\\
15	2.58496250072114\\
15.5	2.58496250072115\\
16	2.58496250072116\\
16.5	2.58496250072116\\
17	2.58496250072116\\
17.5	2.58496250072116\\
18	2.58496250072115\\
18.5	2.58496250072116\\
19	2.58496250072116\\
19.5	2.58496250072115\\
20	2.58496250072115\\
};

\addplot [color=blue, line width=1.0pt, mark=square*,mark repeat=2, mark size = 3pt,mark phase = 2]
  table[row sep=crcr]{%
-6.35133942661603	0.0696056202300159\\
-5.84635386336347	0.086771883450923\\
-5.35478784048414	0.106145231394337\\
-4.84129019908617	0.132126613684122\\
-4.37155966003419	0.163268683303102\\
-3.84930230388465	0.199846940507931\\
-3.34923230465705	0.241523277576411\\
-2.84897483985623	0.295105459220093\\
-2.34905029029878	0.354332891505723\\
-1.84867389708995	0.417074406843195\\
-1.34817495286669	0.495168559012972\\
-0.848641055268479	0.578612916839958\\
-0.348096330936419	0.670980977369927\\
0.150778877295729	0.76036811956943\\
0.65206719718621	0.864554440119151\\
1.15087680277639	0.976429947421855\\
1.65645400192154	1.07885017161918\\
2.15709798289718	1.19315296596505\\
2.65759496985572	1.30067405464675\\
3.15644910842943	1.40914892601762\\
3.6599594480312	1.51415057871583\\
4.18474000904134	1.6095679928176\\
4.70155415208047	1.69768435404251\\
5.21414143910045	1.76973736676161\\
5.72745886268216	1.83247275127181\\
6.24127037629889	1.87856756014395\\
6.7541069145327	1.91487453738762\\
7.26360987600503	1.94294238663746\\
7.77072771985895	1.96440621679186\\
8.27540588467613	1.97684151975229\\
8.77766590200144	1.98790531659096\\
9.27886184815629	1.9938626011072\\
9.77945838681944	1.99698257076001\\
10.2797485040242	1.9980504414134\\
10.7798857024955	1.9992152016152\\
11.2799519895975	1.99983861189839\\
12 2\\
13 2\\
14 2\\
15 2\\
};

\addplot [color=blue, line width=2.0pt, mark=square*,mark repeat=2, mark size = 3pt,mark phase = 2,densely dashed]
  table[row sep=crcr]{%
-5.24038525436822	0.093107936721768\\
-4.74113723837621	0.116091180528885\\
-4.24144069739737	0.143333396657253\\
-3.73964850991061	0.177104588645649\\
-3.23994537170635	0.219039557001089\\
-2.73971300634321	0.263372995924791\\
-2.23968223954248	0.314399998137719\\
-1.73981373938272	0.382033973674342\\
-1.23907855425136	0.450441816354599\\
-0.738637614633179	0.52963296227935\\
-0.237542921766682	0.615406210358954\\
0.262334950925175	0.712081970182248\\
0.762588696654536	0.819580802893962\\
1.26271626633558	0.917946271849633\\
1.76339610028502	1.0296934046648\\
2.26415846277996	1.1454535385664\\
2.76442242830956	1.26563927506961\\
3.26859653775506	1.38572922475196\\
3.77220778676982	1.50538927060779\\
4.27453472966844	1.62224080309899\\
4.7761834021248	1.73127063038027\\
5.28718943883699	1.84730843673003\\
5.79262276380527	1.95700913422761\\
6.30143493606297	2.05454901432082\\
6.81282236783964	2.14147592295556\\
7.35281159217206	2.2251279590137\\
7.86572268650263	2.2930428347705\\
8.37541295366416	2.35339700283995\\
8.88175840737078	2.40609819251948\\
9.38531668536794	2.45603217067718\\
9.88739857740353	2.4927089004608\\
10.3884141745614	2.52282257794984\\
10.8888338049797	2.52619284260172\\
11.3890680681918	2.54078613231745\\
11.8891586718828	2.56139721577101\\
12.3892034254245	2.57488427929296\\
13 2.58\\
13.8 2.58\\
15 2.58\\
};

\begin{pgfonlayer}{foreground}
\draw[<->, thick,line width = 2.5pt, >=stealth,on layer=axis foreground] (axis cs:4,1.6)  node[xshift=-0.2cm,left, fill=white,opacity=0.5,text opacity=1] {\scalebox{1.7}{$\SI{2.2}{dB}$}} -- (axis cs:6.2,1.6);

\draw[<->, thick,line width = 2.5pt, >=stealth,] (axis cs:6.4,2.1)  node[xshift=-0.2cm,left, fill=white,opacity=0.5,text opacity=1] {\scalebox{1.7}{$\SI{2}{dB}$}} -- (axis cs:8.4,2.1);
\end{pgfonlayer}

\end{axis}

\begin{axis}[%
width=7.778in,
height=5.833in,
at={(0in,0in)},
scale only axis,
xmin=0,
xmax=1,
ymin=0,
ymax=1,
axis line style={draw=none},
ticks=none,
axis x line*=bottom,
axis y line*=left
]
\end{axis}
\end{tikzpicture}%
}
    }
    \hspace{0.7cm}
    \subfloat[TD-RC with $L=\SI{2}{km}$.]{
        \scalebox{0.4}{
%
%
%
\begin{tikzpicture}

\pgfdeclarelayer{background}
\pgfdeclarelayer{foreground}
\pgfsetlayers{background,main,foreground}

\begin{axis}[%
width=7in,
height=3.8in,
at={(1.011in,0.642in)},
scale only axis,
xmin=-3.8,
xmax=15,
ymin=0,
ymax=log2(6),
axis background/.style={fill=white},
xtick={-3,-2,...,15},
ytick={0,0.2,...,2.6},
xticklabels={$-3$,,$-1$,,$1$,,$3$,,$5$,,$7$,,$9$,,$11$,,$13$,,$15$},
yticklabels={$0$,,$0.4$,,$0.8$,,$1.2$,,$1.6$,,$2$,,$2.4$,,$2.8$},
xticklabel style={scale=1.8},
yticklabel style={scale=1.8},
xlabel={SNR [dB]},
ylabel={AIR [bpcu]},
xlabel style={scale=2.5},
ylabel style={scale=2.5},
grid=both,
legend pos=south east,
legend style={nodes={scale=2, transform shape}},
]
\addplot [color=red, line width=1.0pt, mark=*,mark repeat=2, mark size = 3pt,mark phase = 2]
  table[row sep=crcr]{%
-15	-0.000628386426122211\\
-14.5	-0.00048420430175374\\
-14	0.000360925340684289\\
-13.5	8.20477017547579e-05\\
-13	-0.00058534711762929\\
-12.5	0.000720103305380082\\
-12	0.000937906121380783\\
-11.5	0.00164558919773842\\
-11	0.00149740918479768\\
-10.5	0.00194930407265725\\
-9.99999999999999	0.00411756933782326\\
-9.50000000000001	0.00501957051473971\\
-8.99999999999999	0.00773916746336393\\
-8.49999999999999	0.00930953698992084\\
-7.99999999999999	0.0112114164872413\\
-7.5	0.0134823144802917\\
-6.99999999999999	0.0163116281104085\\
-6.5	0.0216028190508514\\
-5.99999999999998	0.0257756979505931\\
-5.49999999999999	0.0373779706125592\\
-5	0.0462031782408946\\
-4.50000000000001	0.0560956747940413\\
-3.99999999999999	0.0671368504606763\\
-3.5	0.0850220803520139\\
-3	0.106285842944451\\
-2.49999999999998	0.128981506015365\\
-1.99999999999999	0.162106316837694\\
-1.5	0.199342238128789\\
-0.999999999999987	0.251462065489671\\
-0.499999999999994	0.284861990126885\\
7.71461973242629e-15	0.358935428301359\\
0.5	0.421794137233681\\
1	0.501294565083552\\
1.49999999999999	0.590623786107296\\
1.76091259055682	0.619592770411627\\
2.00000000000002	0.674933496704927\\
2.5	0.763593035430757\\
3.00000000000001	0.86510344566251\\
3.5	0.973931970685943\\
4.00000000000001	1.08006460864625\\
4.5	1.20215902434114\\
5	1.31864372595932\\
5.49999999999998	1.433029442178\\
6.00000000000001	1.5448763147622\\
6.50000000000001	1.6519596907105\\
7.00000000000001	1.74696587331496\\
7.49999999999999	1.82015930789148\\
8	1.88697531822891\\
8.50000000000002	1.9333771349827\\
9	1.96438364795169\\
9.5	1.98187297874564\\
9.99999999999999	1.99038856808567\\
10.5	1.99632251420258\\
11	1.99856080460585\\
11.5	1.99935736863186\\
12	1.99920432422275\\
12.5	1.99979211341982\\
13	1.99995981486682\\
13.5	1.99998140294393\\
14	1.99999017341636\\
14.5	1.99999281580672\\
15	1.99999765503059\\
15.5	1.9999992900057\\
16	1.99999890776879\\
16.5	1.99999972654883\\
17	1.99999960031815\\
17.5	1.9999997216106\\
18	1.99999977241985\\
18.5	1.99999986907066\\
19	1.99999986787566\\
19.5	1.99999986763343\\
20	1.99999988461669\\
};

\addplot [color=red, line width=1.0pt, mark=*,mark repeat=2, mark size = 3pt,mark phase = 2,densely dashed]
  table[row sep=crcr]{%
-15	3.33888990870316e-06\\
-14.5	-8.15256841132002e-06\\
-14	-0.000774787723792988\\
-13.5	-0.000613527652336169\\
-13	-0.000312182759002244\\
-12.5	-0.000480485292578899\\
-12	0.000807410556818174\\
-11.5	0.000746528522666286\\
-11	0.00177805846127921\\
-10.5	0.0024684401265654\\
-9.99999999999999	0.00193077961396892\\
-9.5	0.00329839982250735\\
-8.99999999999999	0.0060321896632638\\
-8.5	0.00766963618850121\\
-7.99999999999998	0.0100694462579534\\
-7.49999999999998	0.0110452205080046\\
-7	0.0166632246869138\\
-6.49999999999998	0.0166392739582616\\
-6	0.0258517574712258\\
-5.49999999999999	0.0301088396647666\\
-4.99999999999999	0.0383580883790253\\
-4.5	0.0470519551595478\\
-4	0.0583139073750642\\
-3.49999999999999	0.0710518291740761\\
-3	0.0944192692957809\\
-2.50000000000001	0.120944270948152\\
-1.99999999999999	0.140517765974492\\
-1.49999999999999	0.175440014043316\\
-0.999999999999997	0.222233742645596\\
-0.500000000000001	0.268412679335959\\
0	0.324044419540866\\
0.50000000000001	0.3807830785822\\
1	0.461512443234244\\
1.5	0.541782575918762\\
1.76091259055681	0.591105240595941\\
2.00000000000001	0.624848229050486\\
2.5	0.721276560508639\\
3	0.817921221859695\\
3.50000000000001	0.924600027668471\\
3.99999999999999	1.03207159275169\\
4.50000000000001	1.15590597451275\\
5.00000000000001	1.28436867782851\\
5.50000000000001	1.39124355459634\\
6.00000000000002	1.53013463880977\\
6.5	1.63582200072114\\
7.00000000000001	1.78113044198566\\
7.50000000000001	1.88932245679165\\
8	2.02261987626973\\
8.5	2.12122967833556\\
9.00000000000001	2.22582532965017\\
9.5	2.31352131372489\\
10	2.39075968653135\\
10.5	2.44966638006049\\
11	2.49708988968901\\
11.5	2.52205352796403\\
12	2.54563362203142\\
12.5	2.55484668390346\\
13	2.56463994068421\\
13.5	2.56806901524012\\
14	2.57120142122346\\
14.5	2.5725867345469\\
15	2.57278132253046\\
15.5	2.57324879964567\\
16	2.57572168015744\\
16.5	2.57660684184832\\
17	2.5774549384163\\
17.5	2.5768780960312\\
18	2.57723926013111\\
18.5	2.5771389424224\\
19	2.57744565888265\\
19.5	2.57813851904386\\
20	2.5791555510438\\
};

\addplot [color=blue, line width=1.0pt, mark=square*,mark repeat=2, mark size = 3pt,mark phase = 2]
  table[row sep=crcr]{%
-6.46316664948084	0.0758418321534544\\
-5.96289733151389	0.0983853878759306\\
-5.46376666603519	0.118457870914653\\
-4.96181002744447	0.147200946805827\\
-4.46241316827945	0.180710468901696\\
-3.96398435034358	0.222269443861335\\
-3.46307341368334	0.267292356314758\\
-2.96086118013505	0.322249377520628\\
-2.4600023895426	0.388981117906582\\
-1.95869217497341	0.458832048307813\\
-1.45678750017605	0.537089075610037\\
-0.953584977546054	0.628464453163292\\
-0.452280126638148	0.72226678695341\\
0.0509982964650568	0.821634689703718\\
0.556461527456859	0.930230320591094\\
1.04578243836289	1.03873424108917\\
1.54533456411707	1.1530099011247\\
2.04349151451383	1.27097115993729\\
2.5564581361032	1.38510822111202\\
3.0760352882081	1.49778817287735\\
3.59989279437328	1.60426355850531\\
4.12541213581592	1.70743016490651\\
4.64578244347124	1.78470724724144\\
5.16993265367468	1.85409719068105\\
5.69145820755736	1.90920454760051\\
6.2124014971205	1.94614165025829\\
6.730703697872	1.97111947734661\\
7.24212006466983	1.98578654738028\\
7.75132388612279	1.99443077907561\\
8.25679805062846	1.99823505984614\\
8.76121599858545	1.99953126025514\\
9.26414331914469	1.99987383380244\\
9.76544759445217	1.99999891794088\\
10.2666137562526	1.999999956874\\
10.7668519623472	1.99999999987224\\
11.266898786404	1.99999999999954\\
11.866898786404	1.99999999999954\\
12.266898786404	1.99999999999954\\
12.666898786404	1.99999999999954\\
13.266898786404	1.99999999999954\\
13.666898786404	1.99999999999954\\
14.266898786404	1.99999999999954\\
14.666898786404	1.99999999999954\\
15.266898786404	1.99999999999954\\
};

\addplot [color=blue, line width=1.0pt, mark=square*,mark repeat=2, mark size = 3pt,mark phase = 2,densely dashed]
  table[row sep=crcr]{%
-5.26030978250317	0.0987838911525723\\
-4.76068865301655	0.123070912795118\\
-4.26071094017503	0.151699076852041\\
-3.759810349612	0.187242594511831\\
-3.2604790284913	0.231148926140708\\
-2.76040046458264	0.277718911777001\\
-2.25982888318559	0.331233309608196\\
-1.75916999220896	0.401492075889333\\
-1.25870329296498	0.472913927096252\\
-0.758226302178605	0.555057524214892\\
-0.257944084555765	0.644156362110082\\
0.242438136373616	0.743926723469307\\
0.742711560217642	0.854977067342673\\
1.24351684142059	0.956698219352735\\
1.74369603748618	1.07297539297214\\
2.24486514776026	1.19335355112318\\
2.75383017091499	1.31926103960412\\
3.25173112926824	1.44484139845809\\
3.75163205860635	1.57107422335096\\
4.25672565746909	1.69665347502972\\
4.77667741586175	1.81673141744516\\
5.27173423130696	1.94267706229912\\
5.77856939370593	2.0649707448901\\
6.28632343188674	2.17219493581032\\
6.79748179697824	2.27045902508306\\
7.31587735403872	2.36065468776028\\
7.84207555969214	2.43123327218235\\
8.35738320037632	2.48782880425223\\
8.86443984346329	2.52788799449346\\
9.37018759261831	2.55663823749215\\
9.87321969606065	2.57220210662777\\
10.3759362313391	2.58029201545813\\
10.8778210472476	2.58363350790954\\
11.3791708344583	2.58455119961842\\
11.8801539485937	2.58492281130547\\
12.3808629700514	2.58496131227105\\
13 2.58096131227105\\
13.5 13 2.58096131227105\\
14 2.58096131227105\\
14.5 13 2.58096131227105\\
15 2.58096131227105\\
};

\begin{pgfonlayer}{foreground}
\draw[<->, thick,line width = 2.5pt, >=stealth,on layer=axis foreground] (axis cs:3.5,1.6)  node[xshift=-0.2cm,left, fill=white,opacity=0.5,text opacity=1] {\scalebox{1.7}{$\SI{2.9}{dB}$}} -- (axis cs:6.4,1.6);

\draw[<->, thick,line width = 2.5pt, >=stealth,] (axis cs:5.8,2.1) node[xshift=-0.2cm,left, fill=white,opacity=0.5,text opacity=1] {\scalebox{1.7}{$\SI{2.6}{dB}$}} -- (axis cs:8.4,2.1) ;
\end{pgfonlayer}

\end{axis}

\begin{axis}[%
width=7.778in,
height=5.833in,
at={(0in,0in)},
scale only axis,
xmin=0,
xmax=1,
ymin=0,
ymax=1,
axis line style={draw=none},
ticks=none,
axis x line*=bottom,
axis y line*=left
]
\end{axis}
\end{tikzpicture}%
}
    }

    \vspace{-1.3cm}
    \subfloat[FD-RC with $L=\SI{5}{km}$.]{
        \scalebox{0.4}{\input{./pics/AIRoverSNR_RC_5.tex}}
    }
    \hspace{0.7cm}
    \subfloat[TD-RC with $L=\SI{5}{km}$.]{
        \scalebox{0.4}{\input{./pics/AIRoverSNR_RC_TD_5.tex}}
    }

    \vspace{-1.3cm}
    \subfloat[FD-RC with $L=\SI{10}{km}$.]{
        \scalebox{0.4}{\input{./pics/AIRoverSNR_RC_10.tex}}
    }
    \hspace{0.7cm}
    \subfloat[TD-RC with $L=\SI{10}{km}$.]{
        \scalebox{0.4}{
%
%
%
\begin{tikzpicture}

\pgfdeclarelayer{background}
\pgfdeclarelayer{foreground}
\pgfsetlayers{background,main,foreground}

\begin{axis}[%
width=7in,
height=3.8in,
at={(1.011in,0.642in)},
scale only axis,
xmin=-3.8,
xmax=20,
ymin=0,
ymax=log2(6),
axis background/.style={fill=white},
xtick={-3,-2,...,20},
ytick={0,0.2,...,2.6},
xticklabels={$-3$,,$-1$,,$1$,,$3$,,$5$,,$7$,,$9$,,$11$,,$13$,,$15$,,$17$,,$19$},
yticklabels={$0$,,$0.4$,,$0.8$,,$1.2$,,$1.6$,,$2$,,$2.4$,,$2.8$},
xticklabel style={scale=1.8},
yticklabel style={scale=1.8},
xlabel={SNR [dB]},
ylabel={AIR [bpcu]},
xlabel style={scale=2.5},
ylabel style={scale=2.5},
grid=both,
legend pos=south east,
legend style={nodes={scale=2, transform shape}}
]

\addplot [color=red, line width=1.0pt, mark=*,mark repeat=2, mark size = 3pt,mark phase = 2]
  table[row sep=crcr]{%
-15	-0.000345789638534229\\
-14.5	-0.00042971611226994\\
-14	-0.000195272304779189\\
-13.5	-0.00145136773724334\\
-13	-0.000614550124921139\\
-12.5	-0.000326439612134966\\
-12	-0.000615684967144592\\
-11.5	-4.61223798886579e-05\\
-11	-0.000417967736367449\\
-10.5	0.00024240481333182\\
-10	0.000490383684727203\\
-9.5	0.000330249651025571\\
-8.99999999999999	0.00150753312058294\\
-8.5	0.000685086591944698\\
-7.99999999999998	0.00188804310835566\\
-7.5	0.00194278836634713\\
-6.99999999999999	0.00280923292431989\\
-6.5	0.00435936717057128\\
-6	0.00426701275996525\\
-5.49999999999999	0.0061564571245117\\
-5	0.00603220845922658\\
-4.5	0.0103695833109036\\
-3.99999999999999	0.01253412217277\\
-3.49999999999999	0.0154505820272985\\
-2.99999999999998	0.0203034912769405\\
-2.49999999999999	0.0247824448970978\\
-2	0.0347441846962286\\
-1.5	0.040567682333251\\
-0.999999999999982	0.048338636660726\\
-0.500000000000007	0.0617396391464347\\
-5.30380106604308e-15	0.0810897752140921\\
0.499999999999988	0.0959236195956598\\
1.00000000000001	0.121857377996477\\
1.50000000000001	0.144351570215259\\
1.76091259055682	0.167975153766605\\
2	0.186601988461564\\
2.5	0.218884446364435\\
3.00000000000001	0.266906582757954\\
3.50000000000001	0.320006204270574\\
4	0.382308719273315\\
4.5	0.436824512556317\\
5.00000000000001	0.517095686075934\\
5.50000000000001	0.594524246061784\\
6.00000000000001	0.674619171972006\\
6.49999999999999	0.76588117788252\\
7.00000000000001	0.848041038780753\\
7.50000000000001	0.931394806715461\\
8.00000000000001	1.01700427964909\\
8.50000000000001	1.0989481459874\\
8.99999999999999	1.1698866120755\\
9.5	1.24653677523586\\
10	1.31456888792106\\
10.5	1.37025496352606\\
11	1.42421535894957\\
11.5	1.46916745738016\\
12	1.51171877225084\\
12.5	1.54777758844206\\
13	1.57112991927088\\
13.5	1.59158917114819\\
14	1.60970979054683\\
14.5	1.62427667188642\\
15	1.63599247099466\\
15.5	1.64735597984176\\
16	1.64696725849113\\
16.5	1.65885911785414\\
17	1.67186890845249\\
17.5	1.66727471265332\\
18	1.6688082567979\\
18.5	1.67008738308046\\
19	1.67458947145598\\
19.5	1.68317035272264\\
20	1.67809905376349\\
};

\addplot [color=red, line width=1.0pt, mark=*,mark repeat=2, mark size = 3pt,mark phase = 2,densely dashed]
  table[row sep=crcr]{%
-15	-0.00121317583770623\\
-14.5	-0.0010797676991429\\
-14	-0.000848878645602754\\
-13.5	-0.000490838243125236\\
-13	-0.00105001771061867\\
-12.5	-0.00108136584167558\\
-12	-0.00132961087371071\\
-11.5	-0.00035827996181285\\
-11	-0.000954222760819246\\
-10.5	-0.000359042543355651\\
-10	-0.000257896586045803\\
-9.5	3.111464939994e-05\\
-8.99999999999999	-0.000239754909047078\\
-8.49999999999999	0.000592334796801097\\
-7.99999999999999	-0.000386160550483942\\
-7.5	0.00113748921200818\\
-7.00000000000001	0.00225683875716184\\
-6.50000000000001	0.00123645142294985\\
-6	0.0034075263762056\\
-5.49999999999999	0.00415876728846139\\
-5	0.00423022485039596\\
-4.49999999999999	0.00891073866932013\\
-3.99999999999999	0.00806319497980172\\
-3.49999999999999	0.0120675326688512\\
-2.99999999999999	0.0133877024195742\\
-2.5	0.0179758742491664\\
-1.99999999999999	0.0265200709988343\\
-1.5	0.0335315668751856\\
-0.999999999999998	0.0425141420962537\\
-0.499999999999997	0.048763599336984\\
-1.44649119982993e-15	0.0641796546325272\\
0.500000000000001	0.0834124248618049\\
1	0.101831307689978\\
1.5	0.12463724738169\\
1.76091259055681	0.139618485970173\\
2	0.148851717504542\\
2.50000000000001	0.186999891827468\\
3	0.225643242643605\\
3.5	0.274512243256734\\
4	0.332779173115277\\
4.5	0.397941839272044\\
5	0.450814007955453\\
5.5	0.529306836591846\\
6	0.60736863027412\\
6.5	0.687429966968762\\
7.00000000000001	0.77874631659803\\
7.5	0.857626031354928\\
8	0.95163258435989\\
8.50000000000001	1.04618055364524\\
9.00000000000002	1.12880943157264\\
9.50000000000001	1.20418909940232\\
10	1.27818464894473\\
10.5	1.3624332834099\\
11	1.41861715356877\\
11.5	1.47848328730475\\
12	1.51830759376575\\
12.5	1.56060328314696\\
13	1.60176194638903\\
13.5	1.63111292230082\\
14	1.65503253984254\\
14.5	1.67351298514243\\
15	1.6917720847056\\
15.5	1.70796123693001\\
16	1.70591795832013\\
16.5	1.73460434305973\\
17	1.73879109723588\\
17.5	1.73464801116399\\
18	1.75423506090232\\
18.5	1.7484445052275\\
19	1.75631694658399\\
19.5	1.75415512525252\\
20	1.7556910427893\\
};

\addplot [color=blue, line width=1.0pt, mark=square*,mark repeat=2, mark size = 3pt,mark phase = 2]
  table[row sep=crcr]{%
-4.73245735130143	0.0144502160465468\\
-4.23067135732117	0.0184616219241377\\
-3.72977138387698	0.0229316713172011\\
-3.22678952382592	0.0282578716316751\\
-2.72073519962525	0.0371327185172593\\
-2.22068559522029	0.0467809154832765\\
-1.70966675490715	0.0580248846475014\\
-1.20417447045663	0.0712965824392901\\
-0.700970361972162	0.0873022583071637\\
-0.193522390469777	0.108350470915493\\
0.314736898099069	0.130729396559168\\
0.823575980043721	0.159082417244661\\
1.33259373440367	0.191417387511092\\
1.84113618035435	0.227584833292083\\
2.34928618881929	0.269861893801931\\
2.8562272993301	0.317022471775578\\
3.36182137998585	0.365910496020892\\
3.86607978877444	0.424378071917726\\
4.36924151922281	0.478017478588996\\
4.87145202128928	0.542254588736627\\
5.37294311748269	0.611170072711306\\
5.87385707636191	0.674052000944069\\
6.37447904338819	0.743922109893358\\
6.87486598094159	0.818171388301443\\
7.37502687484492	0.897907487983596\\
7.87507928071539	0.977157662419788\\
8.37505763495224	1.05527155695608\\
8.87501799329451	1.14578453650949\\
9.37490139167204	1.23182412653733\\
9.49620888216782	1.25875603211782\\
9.99626695241754	1.33997964350914\\
10.4962725386793	1.43078339127753\\
10.9962568080549	1.50738433042433\\
11.4962244520471	1.56558990047484\\
11.9962369065411	1.62744489340304\\
12.4962757605908	1.67228115639695\\
12.9962354838035	1.70855494233378\\
13.4962638615604	1.73374277789253\\
13.9962504124432	1.76541860379727\\
14.4962221283252	1.77230671721679\\
14.9962771179046	1.79807749553801\\
15.4962125439108	1.80970221989292\\
15.9961620353938	1.81144245122046\\
16.4962385405145	1.82365284857335\\
16.9962381029198	1.83051359997966\\
17.4963000759017	1.83547815609112\\
17.9961996875219	1.83171428881846\\
18.4962159957424	1.84247137823144\\
18.9962400209984	1.84145572960667\\
19.4962678033418	1.84482705106273\\
19.9962242653427	1.84506522956839\\
};

\addplot [color=blue, line width=1.0pt, mark=square*,mark repeat=2, mark size = 3pt,mark phase = 2,densely dashed]
  table[row sep=crcr]{%
-3.61858177361006	0.0192249495816609\\
-3.13856165693431	0.0232662673675846\\
-2.61607538498521	0.0314133485065954\\
-2.11199436104833	0.0401625960398077\\
-1.60715983802705	0.047461847683453\\
-1.10128699587942	0.0599752746299073\\
-0.594864531363091	0.0747935642896727\\
-0.0875371161150624	0.0922167386575021\\
0.420705609607929	0.114750218344149\\
0.929781349632047	0.137030886408594\\
1.43918685739566	0.169163746404897\\
1.94828488383109	0.205918960053786\\
2.45691543699811	0.243095088958343\\
2.96445551233683	0.292092391212389\\
3.47056025032973	0.336405781916132\\
3.9752920901131	0.387440147884764\\
4.47878146499532	0.445577198364327\\
4.9811435102777	0.505949855889206\\
5.48270930510132	0.571215904890264\\
5.9838266391135	0.636179632795319\\
6.4843955057485	0.707062929487528\\
6.98478336701511	0.779279398698454\\
7.48497866747008	0.851650611357578\\
7.98504869285535	0.940358643390403\\
8.48505710774519	1.02262021328547\\
8.99679883392245	1.09844387230072\\
9.4968100456598	1.20724246552638\\
9.9968158934635	1.30650828354818\\
10.4968131528227	1.40834595884984\\
10.996817918723	1.51984884270022\\
11.4967875212751	1.60444053075011\\
11.9968479652004	1.70403605354089\\
12.4967726791955	1.77839220658734\\
12.9968090669097	1.85399490944846\\
13.4968279640757	1.93298533948115\\
13.9968550206117	1.9899436837355\\
14.4968103372479	2.02227560049556\\
14.9968331004309	2.07155549298367\\
15.4968516798993	2.10214573683774\\
15.9968329626101	2.1391493609182\\
16.4968119485436	2.15639023846243\\
16.9967973137853	2.15846397984331\\
17.4968287826097	2.176151514284\\
17.9968495576628	2.18625650942308\\
18.4968320462828	2.19596704666083\\
18.9968435172245	2.21682207230177\\
19.4967867606266	2.20235338983251\\
19.9968494659376	2.20958786819161\\
};

\begin{pgfonlayer}{foreground}
\draw[<->, thick,line width = 2.5pt, >=stealth,] (axis cs:19,1.75)  node[left,fill=white,opacity=0.5,text opacity=1,yshift=0.9cm,xshift=-0.2cm] {\scalebox{1.7}{$\SI{0.6}{bpcu}$}} -- (axis cs:19, 2.25);

\draw[<->, thick,line width = 2.5pt, >=stealth,] (axis cs:16,1.63) node[yshift=-0.2cm,below, fill=white,opacity=0.5,text opacity=1] {\scalebox{1.7}{$\SI{0.2}{bpcu}$}}  -- (axis cs:16,1.83);
\end{pgfonlayer}

\end{axis}

\begin{axis}[%
width=7.778in,
height=5.833in,
at={(0in,0in)},
scale only axis,
xmin=0,
xmax=1,
ymin=0,
ymax=1,
axis line style={draw=none},
ticks=none,
axis x line*=bottom,
axis y line*=left
]
\end{axis}
\end{tikzpicture}%
}
    }

    \caption{AIR for FD-RC (left column), TD-RC (right column) and $L=\{1,2,5,10\}\SI{}{km}$.}
    \label{fig:simResults}
\end{figure*}

The numerical simulations were carried out at a carrier  wavelength of $\lambda_c = \SI{1550}{nm}$ with dispersion coefficient $D = \SI{17}{\frac{ps}{nm\cdot km}}$ and attenuation factor $\alpha_\mathrm{SMF} = \SI{0.2}{\frac{dB}{km}}$. The Kerr nonlinearity is neglected.
To modulate the signal, \{4,6\}-PAM modulation is used with  alphabets  $\mathcal{A}_4=\{0,1,2,3\}$ and $\mathcal{A}_6=\{0,1,2,3,4,5\}$, respectively. We choose a symbol rate of $\SI{33}{GBaud}$, and use the predistortion in Fig.~\ref{fig:symbPred}. 
The simulation oversampling factor is $N_\mathrm{os}=50$ for $W\leq 2$ ensuring that the approximation~\eqref{eq:xk_approx} has a negligible error.
The transmitter is bandlimited, i.e., we set $F=2$ which results in $B_\text{tx} = 3 B$; see below~\eqref{eq:filtered_pm_spectrum}. 
The average optical power launched into the fiber is 
\begin{align}
    P_\mathrm{tx} = \lim_{T\to\infty}\,\frac{1}{T}\, \E\Big[||X(t)||^2\Big].
\end{align}
The receive filter $g(t)$ models two low-pass filters, i.e., $g(t) = (g_\mathrm{PD}*g_\mathrm{N})(t)$, where $g_\mathrm{PD}(t)$ is the electrical bandwidth limitation of the PD with one-sided bandwidth $B_\mathrm{PD}/2 = \SI{50}{GHz}$ and the brickwall filter $g_\mathrm{N}(t)$ is adapted to the used pulse shape and passes $99\,\%$ of the energy of $Z''(t)$.  
We measure the noise power $B_\mathrm{PD}\cdot N_0/2$ directly after the PD filter, and normalize $B_\mathrm{PD}\cdot N_0/2 = 1$. Thus, $\mathrm{SNR} = P_\mathrm{tx}$.
Finally, an analog-to-digital converter (ADC) samples at symbol rate $B$.

\begin{figure}[H]
    \scalebox{0.4}{
%
%
\definecolor{mycolor1}{rgb}{0.00000,0.44700,0.74100}%
\definecolor{mycolor2}{rgb}{0.85000,0.32500,0.09800}%
\definecolor{mycolor3}{rgb}{0.92900,0.69400,0.12500}%
\begin{tikzpicture}

\begin{axis}[%
width=7in,
height=4.5in,
at={(1.011in,0.642in)},
scale only axis,
xmin=-2,
xmax=0.5,
ymin=-3.5,
ymax=-2,
axis background/.style={fill=white},
xtick={-2,-1.5,-0.5,0,0.5},
xticklabel style={scale=1.8},
ytick={-3.5,-3,-2.5,-2},
yticklabel style={scale=1.8},
yticklabels={},
tick style={very thick},
xlabel={$\eta_1$},
ylabel={MSE},
xlabel style={scale=2.5},
ylabel style={scale=2.5},
grid=both,
]

\addplot [color=blue, line width=2.5pt, forget plot]
  table[row sep=crcr]{%
-2	-2.59919659557359\\
-1.89655172413793	-2.75135855695339\\
-1.79310344827586	-2.86539214002854\\
-1.68965517241379	-2.95169420576388\\
-1.58620689655172	-3.01774428654919\\
-1.48275862068966	-3.06878520371907\\
-1.37931034482759	-3.10844409208133\\
-1.27586206896552	-3.13919866809836\\
-1.17241379310345	-3.16269822785148\\
-1.06896551724138	-3.17997082115172\\
-0.96551724137931	-3.19154411126489\\
-0.862068965517241	-3.19749724400531\\
-0.758620689655172	-3.19744998142001\\
-0.655172413793103	-3.19048362642467\\
-0.551724137931034	-3.17497409961185\\
-0.448275862068966	-3.14829807925796\\
-0.344827586206897	-3.10634515538639\\
-0.241379310344828	-3.04273198655214\\
-0.137931034482759	-2.94758391797624\\
-0.0344827586206897	-2.80580423894882\\
0.0689655172413794	-2.59518055674644\\
0.172413793103448	-2.28632833506561\\
0.275862068965517	-1.85094460242702\\
0.379310344827586	-1.29102209058256\\
0.482758620689655	-0.68965156726031\\
0.586206896551724	-0.217740159635142\\
};

\addplot [color=red, line width=2.5pt, forget plot, smooth]
  table[row sep=crcr]{%
-2	-3.19416316784592\\
-1.89655172413793	-3.23205416616955\\
-1.79310344827586	-3.26230808049378\\
-1.68965517241379	-3.28727226378229\\
-1.58620689655172	-3.30845617579504\\
-1.48275862068966	-3.32686824328823\\
-1.37931034482759	-3.34320499813497\\
-1.27586206896552	-3.35796146140854\\
-1.17241379310345	-3.37149760103192\\
-1.06896551724138	-3.38407909232786\\
-0.96551724137931	-3.39590211409637\\
-0.862068965517241	-3.40710724519392\\
-0.758620689655172	-3.41778460907531\\
-0.655172413793103	-3.42797006364438\\
-0.551724137931034	-3.43762949083311\\
-0.448275862068966	-3.4466238085827\\
-0.344827586206897	-3.45463869066328\\
-0.241379310344828	-3.46104435330424\\
-0.137931034482759	-3.46460761417796\\
-0.0344827586206897	-3.45287146534131\\
0.0689655172413794	-3.43073249312899\\
0.172413793103448	-3.39692705627761\\
0.275862068965517	-3.33460582975045\\
0.379310344827586	-3.13420788547897\\
0.482758620689655	-2.62880859741926\\
0.586206896551724	-1.45329667933925\\
};

\draw[dashed, black] (axis cs:0,-2) -- (axis cs:0,-3.5);

\addplot[
    only marks, 
    mark=*,
    mark size=5pt, 
    color=blue,
] coordinates {(-0.86, -3.1975)};

\addplot[
    only marks, 
    mark=*,
    mark size=5pt, 
    color=red
] coordinates {(-0.1379, -3.4646)};

\draw[->,>=stealth,line width=2.0pt, black,>={Stealth[scale=1.2]}]
     (axis cs:-0.8,-2.9) .. controls  (axis cs:-0.4,-3.3) .. (axis cs:0.48,-3.45)
    node[pos=0, above, fill=white, inner sep=2.5pt,yshift = 1.5em] {\scalebox{2.5}{increasing SNR}};
     \end{axis}
\end{tikzpicture}%
    }
\caption{MSE at \SI{5}{km} for low-  and high-SNR as a function of $\eta_1$ ($W=1$) using FD-RC pulse shaping. The respective minimum is marked with \addlegendimageintext{mark=*,draw=black,mark size=1.8pt}.}\label{fig:costfct2}
\end{figure}
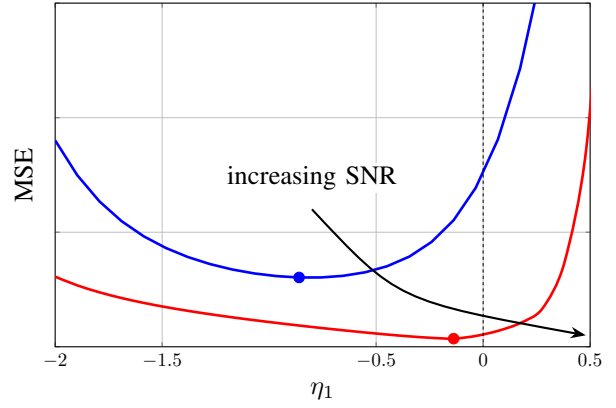

To determine the MSE-optimal coefficients $\boldsymbol{\eta}^\star$ for the fiber length $L$, we use the classic gradient-descent algorithm.%
We find $\boldsymbol{\eta}^\star$ for each SNR. Figure~\ref{fig:costfct2} shows the SNR dependency of the MSE, illustrating that the optimized coefficient $\eta^\star_1$ depends on the SNR. At low SNRs, the MSE is slightly flatter around the minimum, owed to the dominant noise. 
At high-SNR the noise has less impact and chromatic dispersion (CD) takes over as the primary source of distortion. When $\eta_1>0$, the effects of CD are enhanced, leading to a significant increase in the MSE.

\subsection{Numerical Results}
We compute achievable information rates (AIRs)~\cite{garcia2020numerically} as the performance metric. These rates are a lower bound on the mutual information rate, and achievable via mismatched decoding with memoryless Gaussian decoding metrics. Although AIRs are not equivalent to the MSE, one can show via~\cite[Eq.~(46)]{plabst2022achievable} that minimizing the MSE increases the AIRs. 

Figure~\ref{fig:simResults} shows the AIR as a function of SNR for different fiber lengths $L = \{1, 2, 5, 10\}$\SI{}{km}. These fiber lengths are chosen to cover a variety of scenarios, including the impact of selecting $W>1$ and the transition from moderate to severe ISI.
The results are organized into two columns.
The first column corresponds to FD-RC pulse shaping ($\alpha_{\mathrm{ps}} = 0.2$), while the second column corresponds to TD-RC pulse shaping ($\alpha_{\mathrm{ps}} = 0.8$)\footnote{A TD-RC pulse is not bandwidth-limited; however for $\alpha_{\mathrm{ps}} = 0.8$, the 95\% energy bandwidth is slightly larger than $B$, i.e., $~1.2\, B$.}.
The line style indicates the modulation format: Solid lines represent 4-PAM modulation ($M=4$) while dashed lines represent 6-PAM modulation ($M=6$).
We compare the system performance with and without PM. The curves with marker \,\raisebox{0.5\height}{\addlegendimageintext{mark=square*, draw=blue,fill=blue,mark size=1.5pt,only marks}}\, represent results with PM and $W=1$. Additionally, the curves marked with \,\raisebox{0.3\height}{\addlegendimageintext{mark=star, draw=blue,fill=blue,mark size=2.5pt,only marks}}\, represent cases where a second harmonic ($W=2$) has notably improvement compared to $W=1$. 
The marker \,\raisebox{0.5\height}{\addlegendimageintext{mark=*, draw=red,fill=red,mark size=1.5pt,only marks}}\, stands for no PM.
The legend in subplot (a) holds for all plots.

\subsection{Impact of Phase Precoding}
Consider first the curves without PM. For a fixed AIR, the required SNR grows as a function of the fiber length, which increases the total CD. At a certain fiber length, the system becomes interference-limited, where the maximal rate of $\mathrm{log}_2(M)$ bits cannot be achieved regardless of the SNR; see (g)-(h).
Introducing phase precoding (\,\raisebox{0.5\height}{\addlegendimageintext{mark=square*, draw=blue,fill=blue,mark size=1.5pt,only marks}}\,, \,\raisebox{0.3\height}{\addlegendimageintext{mark=star, draw=blue,fill=blue,mark size=2.5pt,only marks}}\,) improves the AIRs by mitigating CD, thereby reducing the required SNR to achieve a specific AIR. This can be seen in the plots (a)-(f) across both modulation formats. For instance, in plot (d), for $L=\SI{2}{km}$, phase precoding results in an SNR gain of $\approx\SI{3}{dB}$ for 4-PAM (\raisebox{0.4\height}{\addlegendimageintext{mark=square*, draw=blue,fill=blue,mark size=1.5pt}}) and $\approx\SI{2.6}{dB}$ for 6-PAM (\raisebox{0.4\height}{\addlegendimageintext{mark=square*, draw=blue,fill=blue,mark size=1.5pt,dashed,dash pattern=on 1pt,}}). Optimizing two coefficients $(\eta_1,\eta_2)$ ($W=2$: \,\raisebox{0.3\height}{\addlegendimageintext{mark=star, draw=blue,fill=blue,mark size=2.5pt,only marks}}\,) gains an additional $\approx\SI{0.6}{dB}$ over $W=1$. However, with increasing fiber length this additional SNR gain decreases. 
For $L\geq \SI{5}{km}$, CD becomes stronger and causes longer inter-symbol interference. 
For $L=\SI{10}{km}$ the system is interference-limited since the PM cannot compensate the CD that extends over several symbol periods; see plot (g) and (h). Here, FD-RC pulses show small AIR improvements in the high-SNR regime. For TD-RC and 6-PAM (\raisebox{0.4\height}{\addlegendimageintext{mark=square*, draw=blue,fill=blue,mark size=1.5pt,dashed,dash pattern=on 1pt,}}), using a PM increases AIRs by $\SI{0.6}{bpcu}$ at high SNRs.

\subsection{Limitations}
The AIR gains of the PM are limited to moderate CD, or equivalently, moderate fiber lengths when operating at a fixed symbol rate. There are two reasons for this: First, the PM precodes phase only, but CD affects both phase and amplitude. Second, increasing CD causes longer inter-symbol interference, which cannot be compensated by a PM that precodes over a single symbol time slot $T_\text{s}$ only. 

\section{Conclusion}
We analyzed a low-cost PM for DD links. To optimize the phase modulator, we used a Fourier series representation of the periodic phase and adapted the closed-form expression from~\cite{plabst2020wiener} for the mean-squared error (MSE) cost function. Numerical results show that the proposed scheme compensates CD for moderate fiber lengths, thus reducing the required transmit power. SNR gains of up to $3\,\text{dB}$ are achieved when using $4$-PAM modulation over $L=2\,\text{km}$.

Future work may explore phase precoding over multiple symbol slots. 
Additionally, adding a simple digital receive filter may further increase AIRs while keeping computational cost low.

\section*{Acknowledgments}
This work was carried out in the framework of the CELTIC NEXT Flagship project AI-NET PROTECT and was supported by the German Federal Ministry of Education and Research under the funding code FKZ16KIS1282.

{\appendix
The gradient of~\eqref{eq:costfunctionU} is 
\begin{align}\label{eq:gradientU}
\frac{\partial\,\epsilon}{\partial \boldsymbol{\eta}} =-2\,\xi^\star\, \frac{\partial}{\partial \boldsymbol{\eta}} \E\left[S_\kappa \, Y_\kappa\right] +\xi^{\star\,2} \, \frac{\partial}{\partial \boldsymbol{\eta}} \E\left[Y_\kappa^2\right],
\end{align}
with the terms

\begin{align}
&\frac{\partial}{\partial \boldsymbol{\eta}} \E\left[S_\kappa \  Y_\kappa\right]=\\
\nonumber
&\,2\,\mu_S^2\,\mathbf{g}^\mathsf{T}\,\mathrm{Re}\big\{\mathrm{j}\big[\mathbf{F}(\mathbf{u}\circ \boldsymbol{\Gamma})\big]\circ\mathbf{w}^\mathsf{*}\big\} +\frac{\sigma_S^2}{2}\,\mathbf{g}^\mathsf{T}\,\mathrm{Re}\big\{\mathrm{j}(\mathbf{F}\circ \mathbf{A})\,\boldsymbol{\Gamma}\big\}\\
\nonumber
&+\sigma_S^2\,\mathbf{g}^\mathsf{T}\,\mathrm{Re}\big\{\mathrm{j}\big[\mathbf{F}(\mathbf{v}\circ \boldsymbol{\Gamma})\big]\circ\mathbf{w}^\mathsf{*}\mkern-5mu-\mathrm{j}\big[\mathbf{F}^\mathsf{*}(\mathbf{u}^\mathsf{*}\mkern-3mu\circ \mkern-3mu\boldsymbol{\Gamma})\big]\mkern-3mu\circ\mkern-3mu(\boldsymbol{\Psi}\, \mathbf{e}_{B_1})\big\}\,,
\end{align}
and 
\begin{align}
    \begin{split}
        &\frac{\partial}{\partial \boldsymbol{\eta}} \E\left[Y_\kappa^2\right]=4\,\mu_S^2\,\mathbf{g}^\mathsf{T}\,\mathrm{Re}\big\{\mathrm{j}\big[\mathbf{F}(\mathbf{u}\circ \boldsymbol{\Gamma})\big]\circ\mathbf{w}^\mathsf{*}\big\}|\,\mathbf{w}|^{\circ 2, \mathsf{T}}\,\mathbf{g}\\
        &+\sigma_S^2\,\mathbf{g}^\mathsf{T}\mathrm{Re}\big\{\mathrm{j}\,\mathrm{diag}(\mathbf{w}^\mathsf{*})\,\mathbf{F}\,\big[(\mathbf{N\,d}) \circ
        \boldsymbol{\Gamma}\big]\\
        & \qquad\qquad\quad-\mathrm{j}\,\mathrm{diag}(\boldsymbol{\Psi} \,\mathbf{ d})\,\mathbf{F}^\mathsf{*}\,(\mathbf{u}^\mathsf{*}\circ \boldsymbol{\Gamma})\big\}\\
        &+\sigma_S^2\,\mathbf{g}^\mathsf{T}\mathrm{Re}\big\{\mathrm{j}\,\mathrm{diag}(\mathbf{w}^\mathsf{*})\,\mathbf{F}\,\big[(\mathbf{N}\,\mathbf{d}^\mathsf{*})\circ\boldsymbol{\Gamma}\big]\\
        & \qquad\qquad\quad-\mathrm{j}\,\mathrm{diag}(\boldsymbol{\Psi}\,\mathbf{d}^\mathsf{*})\,\mathbf{F}^\mathsf{*}\,(\mathbf{u}^\mathsf{*} \circ \boldsymbol{\Gamma})\big\}\\
        &+\sigma_S^2\,\mathbf{g}^\mathsf{T}\mathrm{Re}\big\{\mathrm{j}\,(\mathbf{F} \mkern-3mu\circ\mkern-3mu \mathbf{A})\,\boldsymbol{\Gamma}\,|\mathbf{w}|^{\circ 2,\mathsf{T}}\mkern-3mu+\mkern-3mu\mathrm{j}\,\big(\big[\mathbf{F}\,(\mathbf{u}\mkern-3mu \circ \mkern-3mu\boldsymbol{\Gamma})\big]\mkern-3mu \circ \mathbf{w}^\mathsf{*}\big)\mathbf{z}^\mathsf{T}\big\}\,\mathbf{g}\\
        &+\frac{\sigma_S^4}{8\,\mu_S^2}\,\mathbf{g}^\mathsf{T}\mathrm{Re}\big\{\mathrm{j}\,\big[\mathbf{F} \mkern-3mu\circ\mkern-3mu(\mathbf{Q}\,\boldsymbol{\Psi}\,\mathbf{N}^\mathsf{T})\big]\,\boldsymbol{\Gamma}+\mathrm{j}\,\big[(\boldsymbol{\Psi}\,\mathbf{N}^\mathsf{T})\mkern-3mu \circ \mkern-3mu(\mathbf{Q}\,\mathbf{F})\big]\boldsymbol{\Gamma}\big\}\\
        &+\frac{\sigma_S^4}{8\,\mu_S^2}\, \mathbf{g}^\mathsf{T}\mathrm{Re}\big\{\mathrm{j}\,\big[\mathbf{F}\mkern-3mu\circ \mkern-3mu(\mathbf{P}\,\boldsymbol{\Psi}^\mathsf{*}\,\mathbf{N}^\mathsf{T})\big]\,\boldsymbol{\Gamma}+\mathrm{j}\,\big[(\boldsymbol{\Psi}\,\mathbf{N}^\mathsf{H}) \mkern-3mu\circ \mkern-3mu(\mathbf{P}\,\mathbf{F}^\mathsf{*})\big]\boldsymbol{\Gamma}\big\}\\ 
        &+\frac{\sigma_S^4}{4\,\mu_S^2}\,\mathbf{g}^\mathsf{T}\mathrm{Re}\big\{\mathrm{j}\,(\mathbf{F} \circ \mathbf{A})\,\boldsymbol{\Gamma}\big\}\,\mathbf{z}^\mathsf{T}\,\mathbf{g}\\
        &+\frac{\mu_{4,S}-3\,\sigma_S^4}{4\,\mu_S^2}\,\mathbf{g}^\mathsf{T}\mathrm{Re}\big\{\mathrm{j}\,(\mathbf{F} \circ \mathbf{B})\,\boldsymbol{\Gamma}\big\},
    \end{split}
\end{align}
and definitions  
\begin{align*}
&\mathbf{A} = \boldsymbol{\Psi}^\mathsf{*}\mathbf{N}^\mathsf{T}, \, 
\mathbf{B} = \big[(\mathbf{g}^\mathsf{T}|\boldsymbol{\Psi}|^{\circ 2})\mkern-3mu \circ \mkern-3mu \boldsymbol{\Psi}^\mathsf{*}\big] \mathbf{N}^\mathsf{T},\, \mathbf{d} = \boldsymbol{\Psi}^\mathsf{T}\mathrm{diag}(\mathbf{w}^\mathsf{*}) \mathbf{g},\\
&\mathbf{N} =\mkern-2mu (\mathbf{H} \circ \exp{(\mathrm{j} \boldsymbol{\varphi}')}), \,
\mathbf{P} = \big(\boldsymbol{\Psi}^\mathsf{*} \boldsymbol{\Psi}^\mathsf{T}\big) \circ \mathbf{g}^\mathsf{T}, \,\\
&\mathbf{Q} = \big(\boldsymbol{\Psi}^\mathsf{*}\, \boldsymbol{\Psi}^\mathsf{H}\big) \circ \mathbf{g}^\mathsf{T}, \,\mathbf{u} = \mathbf{N}\,\mathbf{1}_K, \,
\mathbf{v} = \mathbf{N}\,\mathbf{e}_{B_1},
\end{align*}
$\boldsymbol{\Gamma}\in \mathbb{R}^{M_f\times W}$ corresponding to~\eqref{eq:varphiPrime} with
\begin{align}
     \boldsymbol{\Gamma} &=  [  
     \underbrace{(\boldsymbol{\gamma}_0, \boldsymbol{\gamma}_1 \ldots \boldsymbol{\gamma}_{N_\mathrm{os}-1})}_{\text{Repeat $B_2$ times}},\;
     \boldsymbol{\gamma}_0,\;
     \underbrace{(\boldsymbol{\gamma}_1, \ldots \boldsymbol{\gamma}_{N_\mathrm{os}-1},\boldsymbol{\gamma}_0)}_{\text{Repeat $B_2$ times}}]^\mathsf{T}.
\end{align}
}


\end{document}